\documentclass[final,5p,times,twocolumn,numbers]{elsarticle}

\usepackage{amssymb}
\usepackage{amsfonts}
\usepackage{amsmath}
\usepackage{ulem}
\usepackage{cases,bm}
\usepackage{graphicx}
\usepackage[colorlinks=true, allcolors=blue]{hyperref}
\usepackage{cleveref}
\newcommand{\pd}{\partial}
\newcommand{\dd}{\mathrm{d}}
\newcommand{\ee}{\mathrm{e}}
\newcommand{\ii}{\mathrm{i}}

\begin{document}
\begin{frontmatter}
\title{Kink collisions in a two-dimensional gravity model}

\author[address1,address2]{Zhen-Tao He}

\author[address1]{Yuan Zhong\corref{mycorrespondingauthor}}
\cortext[mycorrespondingauthor]{Corresponding author}
\ead{zhongy@mail.xjtu.edu.cn}

\address[address1]{MOE Key Laboratory for Nonequilibrium Synthesis and Modulation of Condensed Matter, \\ School of Physics, Xi’an Jiaotong University, Xi’an 710049, China}
\address[address2]{School of Physical Sciences, University of Chinese Academy of Sciences, Beijing 100049, China}
\date{\today}

\begin{abstract}
We numerically study kink-antikink collisions in the self-gravitating $\phi^4$ model coupled to the two-dimensional dilaton gravity theory proposed by Mann et al. The static kink solutions interpolate between an anti-de Sitter (AdS$_2$) region and a Minkowski region, and can be regarded as two-dimensional analogues of certain thick branes. By scanning the initial velocity for several gravitational couplings, we find that gravity modifies the scattering structure: the resonance windows shift toward higher initial velocities and become progressively narrower as the coupling $\kappa$ increases, while the critical escape velocity increases mildly. A linear perturbation analysis further indicates that the shape modes turn into long-lived quasi-bound states in the weak-gravity regime, which could leak energy during collisions and may therefore contribute to the shift and narrowing of the windows and the increase of the critical velocity. The collisions further produce a clear geometrical response: the conformal factor decreases after the collision, corresponding to a contraction of the local proper spatial scale in the conformal gauge, and this effect becomes stronger for larger $\kappa$. Meanwhile, the Ricci scalar develops transient peaks during kink encounters but remains finite in all simulations considered. Thus, in contrast to higher-dimensional thick-brane collisions, we find no evidence for spacetime singularity formation in this two-dimensional model.
\end{abstract}


\end{frontmatter}

\section{Introduction}
Topological defects arise naturally in systems undergoing phase transitions and spontaneous symmetry breaking, and play an important role in many areas of modern physics, ranging from condensed matter physics to cosmology~\cite{Vachaspati2007,Campo2014}.
For instance, cosmic strings and domain walls produced during cosmological phase transitions in the early universe could offer mechanisms for the formation of large-scale structure~\cite{Kibble1976,Kibble1980,Vilenkin1981}.
Moreover, thick-brane models are smooth generalizations of the Randall-Sundrum thin-brane models~\cite{Randall1999,Randall1999a}. In these models, our four-dimensional (4D) world can be described as a thick brane supported by a kink-like scalar configuration interpolating between two five-dimensional anti-de Sitter (AdS$_5$) regions~\cite{Skenderis:1999mm,DeWolfe:1999cp,Gremm:1999pj}. Bulk matter fields can also be localized on the thick brane; see Refs.~\cite{Dzhunushaliev2010,Liu2018} for reviews.

Interactions among topological defects constitute an essential problem. Even for kinks, perhaps the simplest topological defects, their mutual interactions exhibit abundant and intriguing phenomena. In the celebrated $\phi^4$ model, a kink-antikink pair escapes to infinity when the initial velocity exceeds a critical value, $v_c\simeq0.26$~\cite{Sugiyama1979}. For $v<v_c$, the kink-antikink pair may form a bound state, known as a bion, in which the pair bounces repeatedly after collision; alternatively, it may eventually escape after a finite number of bounces when the initial velocity lies within certain intervals known as fractal resonance windows~\cite{Moshir1981,Campbell1983,Anninos1991,Goodman2007,Dorey2017}. This peculiar phenomenon can be explained by resonant energy exchange between the translational and shape modes of the linear perturbation spectrum~\cite{Campbell1983}.

The well-known resonant energy-exchange mechanism has been successfully applied to resonance windows in a wide range of models, including sine-Gordon-like models~\cite{Campbell1986,Gani1999,Gani2022,Dorey2021,CarreteroGonzalez2022,Hora2025}, polynomial models~\cite{Hoseinmardy2010,Belendryasova2017,Gani2021,Khare2022,MoradiMarjaneh2022,Bazeia2023}, logarithmic models~\cite{Belendryasova2021}, wobbling-kink models~\cite{Campos2021,AlonsoIzquierdo2021,AlonsoIzquierdo2022}, and kink-impurity models~\cite{Kivshar1991,Fei1992,Zhou2017,Lizunova2021}, among others~\cite{Zhong2020,Yan2020,Santos2025}. Nevertheless, exceptions to this mechanism also exist. A striking example is provided by the $\phi^6$ model~\cite{Dorey2011,Gani2014,Yan2022,Adam2022}. Although isolated $\phi^6$ kinks do not possess shape modes, a collective bound mode of the kink-antikink pair can instead mediate the energy exchange. In addition, other ingredients, such as quasinormal modes (QNMs)~\cite{Dorey2018,Campos2020}, fermion fields~\cite{Campos2022}, and delocalized modes~\cite{Santos2025}, may also be responsible for energy exchange. Advances in the collective-coordinate method~\cite{Weigel2014,Takyi2016,Weigel2018,Manton2021,Manton2021a,Adam2022a,Pereira2021,Pereira2023}, together with the discovery of the relation between bounce windows and the separatrix map~\cite{RoyH.2006}, offer a more quantitative understanding of the intricate energy-exchange mechanism.

When coupled to gravity, collisions of domain walls or bubbles can have important cosmological implications~\cite{Hawking1982,BlancoPillado2004,Freivogel2007,Easther2009,Giblin2010,Johnson2012,Hwang2012,Bond2015}; see~\cite{Aurrekoetxea2025} for a review. Collisions of thick branes in asymptotically AdS$_5$ spacetimes may provide a reheating mechanism for ekpyrotic brane-universe scenarios~\cite{Takamizu2006}. Spacelike singularities generally arise from such brane collisions~\cite{Takamizu2006,Takamizu2007,Omotani2011,Maeda2012} unless the energy conditions are violated~\cite{Tziolas2008,Tziolas2009}. Studies of cosmic bubble collisions have established quantitative connections between scalar-field inflationary models and signatures imprinted on the cosmic microwave background (CMB)~\cite{Wainwright2014,Wainwright2014a,Johnson2016,Johnson2016a}. In addition, gravitational waves emitted from bubble collisions can exhibit distinctive spectral signatures~\cite{Kim2015}.

Since kinks are topological soliton solutions in $(1+1)$-dimensional field theories, it is natural to couple them to two-dimensional gravity and study the interplay between nonlinear dynamics and gravitational effects. Two-dimensional gravity provides a simplified framework for studying difficult issues in gravitational theory, such as the quantization of gravity~\cite{Henneaux1985,Alwis1992}, gravitational collapse~\cite{Vaz1994}, and black hole evaporation~\cite{Callan1992,Bilal1993,Russo:1992ht,Russo:1992ax,Thorlacius:1994ip}; see~\cite{doi:10.1142/0622,Nojiri2001,Grumiller:2002nm,Mertens:2022irh} for reviews. In a two-dimensional gravity model proposed by Mann et al.~\cite{Mann1991}, which we shall refer to as MMSS gravity, exact self-gravitating kink solutions without curvature singularities exist~\cite{Stoetzel1995,AlonsoIzquierdo:2019hvl}, and the corresponding metrics describe asymptotically AdS$_2$ geometries. In Ref.~\cite{Zhong2021}, one of the present authors further revealed similarities between two-dimensional self-gravitating kinks and five-dimensional thick-brane models, and extended the superpotential method used in brane-world models~\cite{Skenderis:1999mm,DeWolfe:1999cp,Gremm:1999pj} to MMSS gravity. This provides a systematic approach to constructing exact self-gravitating kink solutions in MMSS gravity~\cite{Zhong:2021voa,Zhong:2022pkv,Feng:2022ndn,Zhong:2023pel,Wang:2024dbf,Wang:2024kxo}.

Then a natural question arises: how do self-gravitating kinks interact with each other? Investigating this question could reveal how kink collisions are affected by two-dimensional gravity, deepen our understanding of two-dimensional gravity, and provide insights into similar issues concerning the collisions of higher-dimensional self-gravitating domain walls. In this work, we study the collision of $\phi^4$ kinks in MMSS gravity. The collision of sine-Gordon kinks will be studied in a forthcoming companion paper.

The paper is organized as follows. In Sec.~\ref{model}, we introduce self-gravitating $\phi^4$ kinks in MMSS gravity and present the basic equations for evolving their collisions. In Sec.~\ref{result}, we show typical numerical results and discuss their physical implications. Finally, in Sec.~\ref{conclusion}, we summarize our conclusions. Throughout this paper, we use natural units.

\section{The model and numerical methods}
\label{model}

\subsection{Static kink solution}
We consider MMSS gravity minimally coupled to a canonical real scalar field $\phi$. The action is given by~\cite{Stoetzel1995,AlonsoIzquierdo:2019hvl,Zhong2021}
\begin{equation}\label{action}
S=\frac{1}{\kappa}\int \dd^2x\sqrt{-g}\left[ -\frac{1}{2}\partial ^{\mu}\varphi \partial _{\mu}\varphi +\varphi R+\kappa \mathcal{L} _{\mathrm{m}} \right],
\end{equation}
where $\varphi$ is the dilaton field, $R$ is the Ricci scalar, and $\kappa>0$ is the gravitational coupling constant. The matter sector is described by the Lagrangian
\begin{equation}
\mathcal{L} _{\mathrm{m}}=-\frac{1}{2}\partial ^{\mu}\phi \partial _{\mu}\phi -V(\phi).
\end{equation}
Here $V(\phi)$ is the scalar potential that supports kink configurations. In the numerical analysis below, the scalar field $\phi$ and the coupling constant $\kappa$ are rescaled to be dimensionless.

For the static metric ansatz
\begin{equation}
\dd s^2=-\ee^{2A(x)}\dd t^2+\dd x^2,
\end{equation}
the dilaton equation admits the special solution
\begin{equation}
\varphi(x)=2A(x),
\end{equation}
and the dynamical equations can be written in a simple first-order form, from which exact gravitating kink solutions can be constructed~\cite{Zhong2021}. As an example, we consider an exactly solvable model with the following scalar potential~\cite{Zhong2021}:
\begin{equation}
V(\phi)=\frac{1}{2d^2}\left( \phi ^2-1 \right) ^2-\frac{\kappa}{8d^2}\left( \phi -\frac{\phi ^3}{3}-\frac{2}{3} \right) ^2,
\end{equation}
where $d$ is a constant parameter.

This potential supports a $\phi^4$-type kink solution:
\begin{eqnarray}
\label{K}
\phi _{\mathrm{K}}\left( x \right)&=&\tanh \left( \frac{x}{d} \right),\\
\label{warpFactorK}
A_{\mathrm{K}}(x)&=&-\frac{\kappa}{6}\left[ \ln\cosh
        \left(\frac{x}{d}\right)
        +\frac{1}{4}\tanh ^2\left(\frac{x}{d}\right)-\frac{x}{d} \right],
\end{eqnarray}
and the corresponding antikink solution:
\begin{eqnarray}
 \phi_{\bar{\mathrm{K}}}(x)&=&\phi_{\mathrm{K}}(-x),\\
 \label{AK}
 A_{\bar{\mathrm{K}}}(x)&=&A_{\mathrm{K}}(-x).
\end{eqnarray}
The parameter $d$ sets the characteristic thickness of the kink and antikink.

For the kink solution, the asymptotic behavior of the metric is
\begin{equation} \label{A_asymptotic}
   A_{\mathrm{K}}=
    \begin{cases}
        kx+A_{\mathrm{M}},\quad & x\rightarrow-\infty
        \\
        A_{\mathrm{M}},\quad  & x\rightarrow+\infty
    \end{cases}
\end{equation}
where $k=\frac{\kappa}{3d}$ and $A_{\mathrm{M}}=\frac{\kappa}{6}(\ln2-\frac{1}{4})$. Therefore, the geometry of the kink solution is asymptotically AdS$_2$ on the left and asymptotically flat on the right. 
For the antikink solution,
\begin{equation}\label{Abar_asymptotic}
A_{\bar{\mathrm{K}}}=
\begin{cases}
A_{\mathrm{M}},\quad & x\rightarrow-\infty
\\
-kx+A_{\mathrm{M}},\quad  & x\rightarrow+\infty
\end{cases}
\end{equation}
and the geometry is asymptotically flat on the left and asymptotically AdS$_2$ on the right.
The kink and antikink solutions have forms similar to those of the five-dimensional thick-brane solutions studied in Ref.~\cite{Takamizu2006}, and may therefore be regarded as two-dimensional analogues of thick branes.

\subsection{Dynamical and constraint equations}    
As in Ref.~\cite{Takamizu2006}, we simulate the collision of self-gravitating kink-antikink pairs in the conformally flat gauge:
\begin{equation}\label{conformal_gauge}
\dd s^2=\ee^{2A(t,z)}\left( -\dd t^2+\dd z^2 \right).
\end{equation}
Under this gauge, we construct approximate moving kink/antikink profiles by boosting the static solutions in Eqs.~\eqref{K}-\eqref{AK}. 

The dilaton equation takes the following form
\begin{equation}\label{wave_A}
\ddot{\varphi}-2\ddot{A}=\varphi ^{\prime\prime}-2A^{\prime\prime},
\end{equation}
where the dot and the prime denote $\partial/\partial t$ and $\partial/\partial z$, respectively.
The dilaton equation has more general solutions, but for simplicity, we only consider the special case with 
\begin{equation}
\varphi(t,z)=2A(t,z).
\end{equation}
This algebraic relation reduces the number of independent unknown variables and simplifies the form of the field equations. 
We then obtain two dynamical equations
\begin{eqnarray}
\label{wave_dilaton}
-\ddot{A}+A ^{\prime\prime}&=&-\frac{\kappa}{2} \ee^{2A} V,\\	
\label{wave_phi}
-\ddot{\phi}+\phi ^{\prime\prime}&=&\ee^{2A}\frac{\dd V}{\dd\phi},
\end{eqnarray}
a momentum constraint
\begin{equation}\label{constraint1}
	\mathcal{M}\equiv 4(\dot{A}^\prime-\dot{A}A^\prime)+\kappa \dot{\phi}\phi ^\prime=0,
\end{equation}
and a Hamiltonian constraint
\begin{equation}\label{constraint2}
	\mathcal{H}\equiv  \dot{A}^2+A^{\prime2}-2A^{\prime\prime} 	
	-\frac{\kappa}{4} \left( 2\ee^{2A}V+\phi ^{\prime2}+\dot{\phi}^2 \right)=0.
\end{equation}
These equations are much simpler than their five-dimensional counterparts~\cite{Takamizu2006}, making the numerical evolution very efficient.

To solve the dynamical equations numerically, we use finite differencing with fourth-order accurate stencils and the built-in \textsf{MATLAB} solver \textsf{ode78} for time integration with adaptive time steps.
Violations of the constraints are used as a check of the numerical accuracy.

In the numerical simulations, all length scales are normalized by the kink thickness $d$.
The remaining free parameters are the gravitational coupling constant $\kappa$ and the initial velocity $v$.

\subsection{Initial data}
We start by constructing a single moving kink $\phi_{\mathrm{K}}(t,z-z_0;v)$, initially centered at $z=z_0$ and moving with velocity $v$ in the $(t,z)$ coordinates of \eqref{conformal_gauge}.
This can be done by a Lorentz boost,
\begin{equation}
    \phi_{\mathrm{K}}(t,z-z_0;v)=
    \phi_{\mathrm{K}}[x(z^*)],
\end{equation}
where $z^* = \gamma(z-z_0-vt)$ and $x$ is implicitly determined by
$z^* = \int \ee^{-A_{\mathrm{K}}(x)}\,\dd x$, with
$\gamma = 1/\sqrt{1-v^2}$.
The warp factor $A_{\mathrm{K}}(t,z-z_0;v)$ is transformed in the same way.

The initial data $\{\phi,\pd_t\phi,A,\pd_tA\}_{t=0}$ for a kink and an antikink separated by a distance $2z_0$ and approaching each other with the same speed $v$ are then approximately constructed by superposing a boosted kink and antikink:
\begin{align} 
    \phi(t,z)
    &=
    \phi_{\mathrm{K}}(t,z+z_0;v)
    +
    \phi_{\bar{\mathrm{K}}}(t,z-z_0;-v)
    -1,\label{phiinit}
    \\
    A(t,z)
    &=
    A_{\mathrm{K}}(t,z+z_0;v)
    +
    A_{\bar{\mathrm{K}}}(t,z-z_0;-v)
    -A_{\mathrm{M}}. \label{Ainit}
\end{align}
In the above construction, the kink is placed on the left and the antikink on the right.
Together with the asymptotic behaviors in Eqs.~\eqref{A_asymptotic} and \eqref{Abar_asymptotic}, this gives the geometric configuration
$\mathrm{AdS}_2$-$\mathrm{Minkowski}$-$\mathrm{AdS}_2$.
The two inward-facing Minkowski tails have $A'\to0$ and can therefore be superposed smoothly.
One may ask whether the order of the kink-antikink pair can be interchanged, with an antikink on the left and a kink on the right, so that the geometric configuration becomes
$\mathrm{Minkowski}$-$\mathrm{AdS}_2$-$\mathrm{Minkowski}$.
For this reversed sector, however, a direct superposition retains unwanted exterior AdS$_2$ tails, while matching the two AdS$_2$ sides to flat exteriors produces a discontinuity in $A'$.
Thus, the reversed sector cannot be constructed using the present linear-superposition prescription.

\begin{figure}
    \centering
    \includegraphics[width=\linewidth]{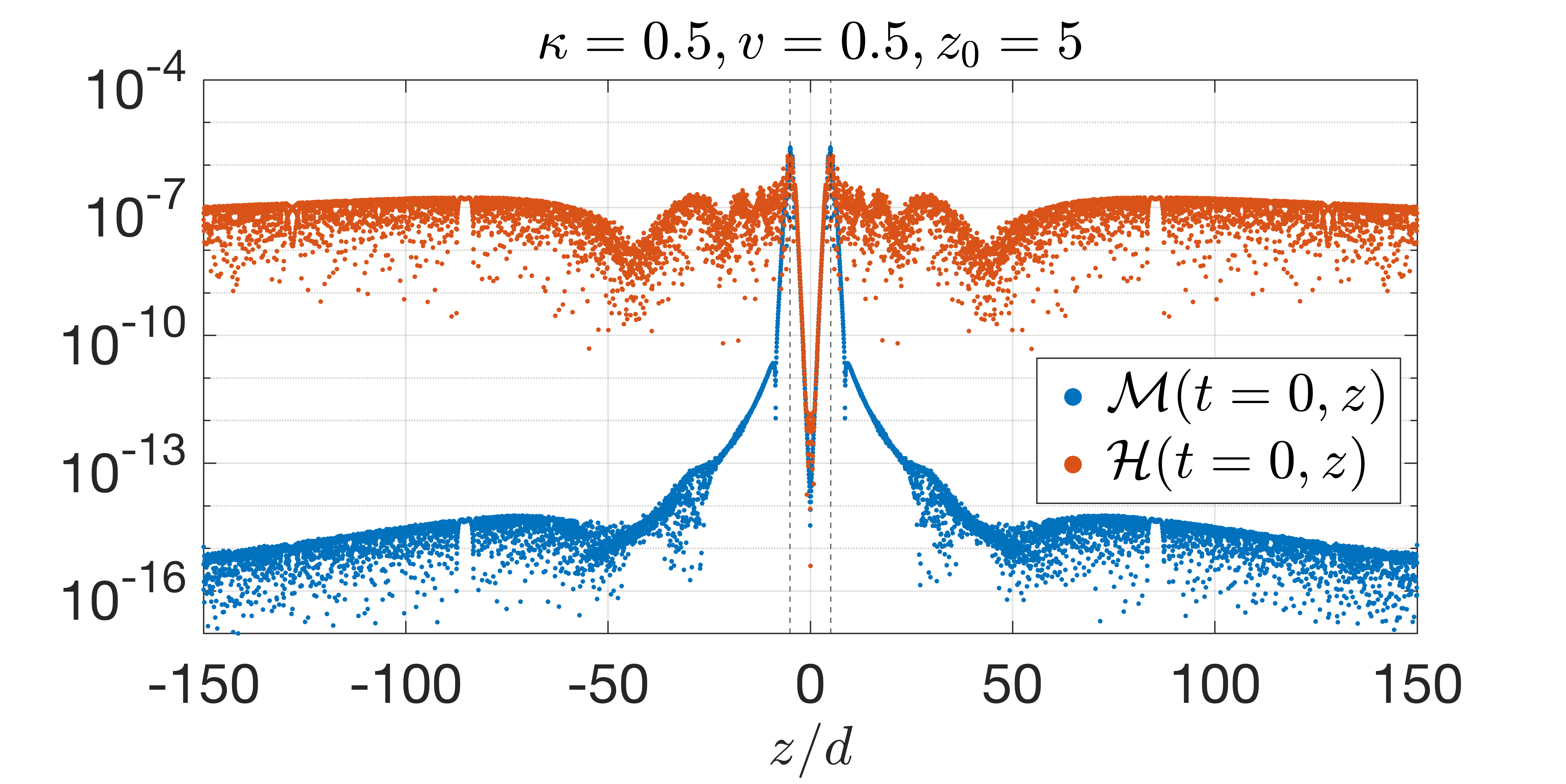}
    \caption{Constraint violations of the initial data constructed by superposing a boosted kink-antikink pair.
    The magnitudes of the violations are typically no larger than $10^{-6}$ over the parameter range $\kappa\leq0.5$ and $v\leq0.5$ considered in this work.
    Larger values of $\kappa$ or $v$ lead to larger localized violations near the centers of the kink and antikink, $z=\pm z_0$.
    These violations can dominate over those induced by the boundary conditions and propagate outward with growing amplitude.
    Here we take $N=7501$ grid points on the domain $z/d\in[-150,150]$.
    }
    \label{c_init}
\end{figure}

As shown in Fig.~\ref{c_init}, this superposition satisfies the constraints \eqref{constraint1} and \eqref{constraint2} well initially.
Although localized spikes in the constraint violations appear near the centers of the kink and antikink, their magnitudes are typically no larger than $10^{-6}$ over the parameter range $\kappa\leq0.5$ and $v\leq0.5$ considered in this work.
We also find that increasing the initial separation $2z_0$ does not significantly reduce the magnitude of these localized spikes, and we therefore take $z_0=5$ throughout this work.

We checked that the initial momentum constraint $\mathcal{M}$ exhibits the expected fourth-order convergence with the number of grid points $N$.
However, the magnitude of the Hamiltonian constraint remains basically unchanged as the number of grid points $N$ increases, indicating a limitation inherent in the superposition approximation \eqref{phiinit} and \eqref{Ainit}.

In principle, the constraint violations in the superposed initial data can be eliminated by solving the constraint equations at the initial time.
The constraint equations \eqref{constraint1} and \eqref{constraint2} may be regarded as a system of ordinary differential equations for $A(t=0,z)$ and $\dot{A}(t=0,z)$, once $\phi(t=0,z)$ and $\dot{\phi}(t=0,z)$ are specified.
Only the initial values of $A(t=0,z)$ and $\dot{A}(t=0,z)$ at one point are then required, corresponding to the remaining gauge freedom.
However, this procedure increases the computational cost in parameter scans, since the coordinate transformation $z\leftrightarrow x$ has no closed analytic form and must be evaluated numerically many times.
We also attempted a Newton iteration using \eqref{Ainit} as an initial guess, but the Jacobian matrix was found to be singular.

\subsection{Boundary conditions}
We truncate the computational domain by introducing artificial timelike boundaries at $z=\pm L$.
Constructing boundary conditions that preserve the constraints exactly during the evolution is, however, nontrivial.
Since we are interested in the dynamics in the collision region, $|z|\lesssim z_0$, we take $L/d\gg1$ and impose boundary conditions based on the asymptotic behavior of the dynamical fields, following the strategy adopted in Ref.~\cite{Takamizu2006}.

Specifically, we impose Dirichlet boundary conditions,
$\phi|_{z=\pm L}=-1$, for the scalar field.
For the warp factor, we impose the Neumann boundary conditions
$A^\prime(z=\pm L)=\mp k\gamma \ee^{A(t,z=\pm L)}$.
The boundary value of $A(t,z=\pm L)$ is determined from the asymptotic form \eqref{A_asymptotic}, transformed into the $z$ coordinate.
At large $|z|$, this gives
\begin{equation}
    \ee^{A(t,z=\pm L)} \simeq
    -\frac{1}{k[\gamma(-L+z_0-vt)+C]},
\end{equation}
where $C$ is a constant fixed by the initial data \eqref{Ainit}.

\begin{figure}
    \centering
    \includegraphics[width=\linewidth]{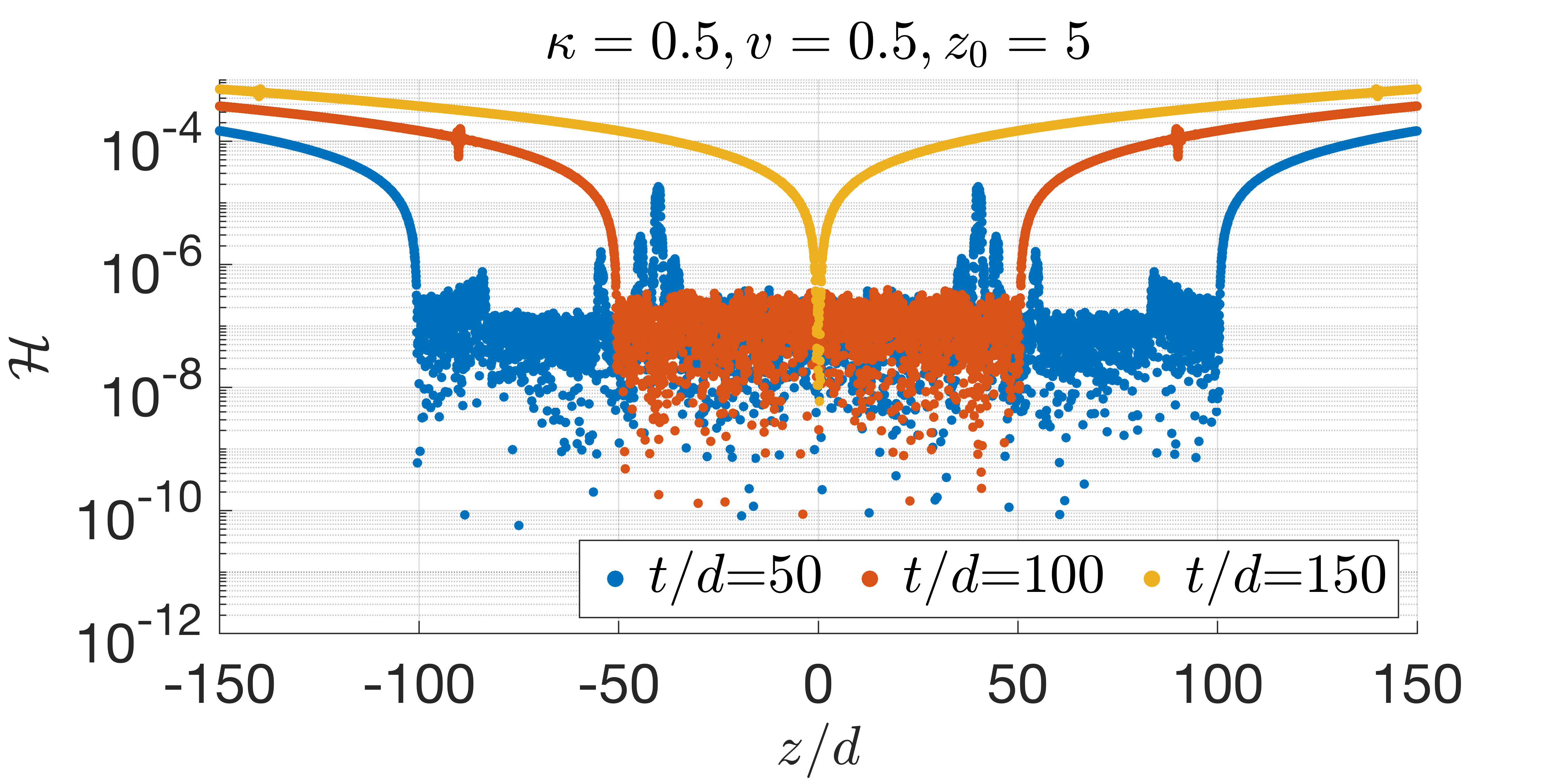}
    \caption{Snapshots of the Hamiltonian constraint $\mathcal{H}$.
    The plot shows that the constraint violations induced by the approximate initial data and by the boundary conditions propagate approximately along null characteristics.
    We have also monitored the momentum constraint $\mathcal{M}$ and found that it exhibits the same qualitative behavior, with smaller violations in the parameter range considered here.
    The violations induced by the initial data become smaller for smaller values of $\kappa$ and $v$.
    }
    \label{Hconstraint}
\end{figure}

As shown in Fig.~\ref{Hconstraint}, the constraint violations induced by the approximate initial data and by the boundary conditions remain below $10^{-3}$ during the evolution.
The violations generated near the artificial boundaries propagate inward approximately along null characteristics.
We therefore restrict the presentation of the numerical results to the region $|z|<L-t$, which lies inside the future domain of dependence of the initial data and is not affected by boundary-generated constraint violations.
Within this region, the constraint violations are much smaller, typically of order $10^{-7}$ apart from early-time localized spikes.

\section{Results}
\label{result}

\subsection{Scattering outcomes and resonance windows}

\begin{figure}[t]
    \centering
    \includegraphics[width=\linewidth]{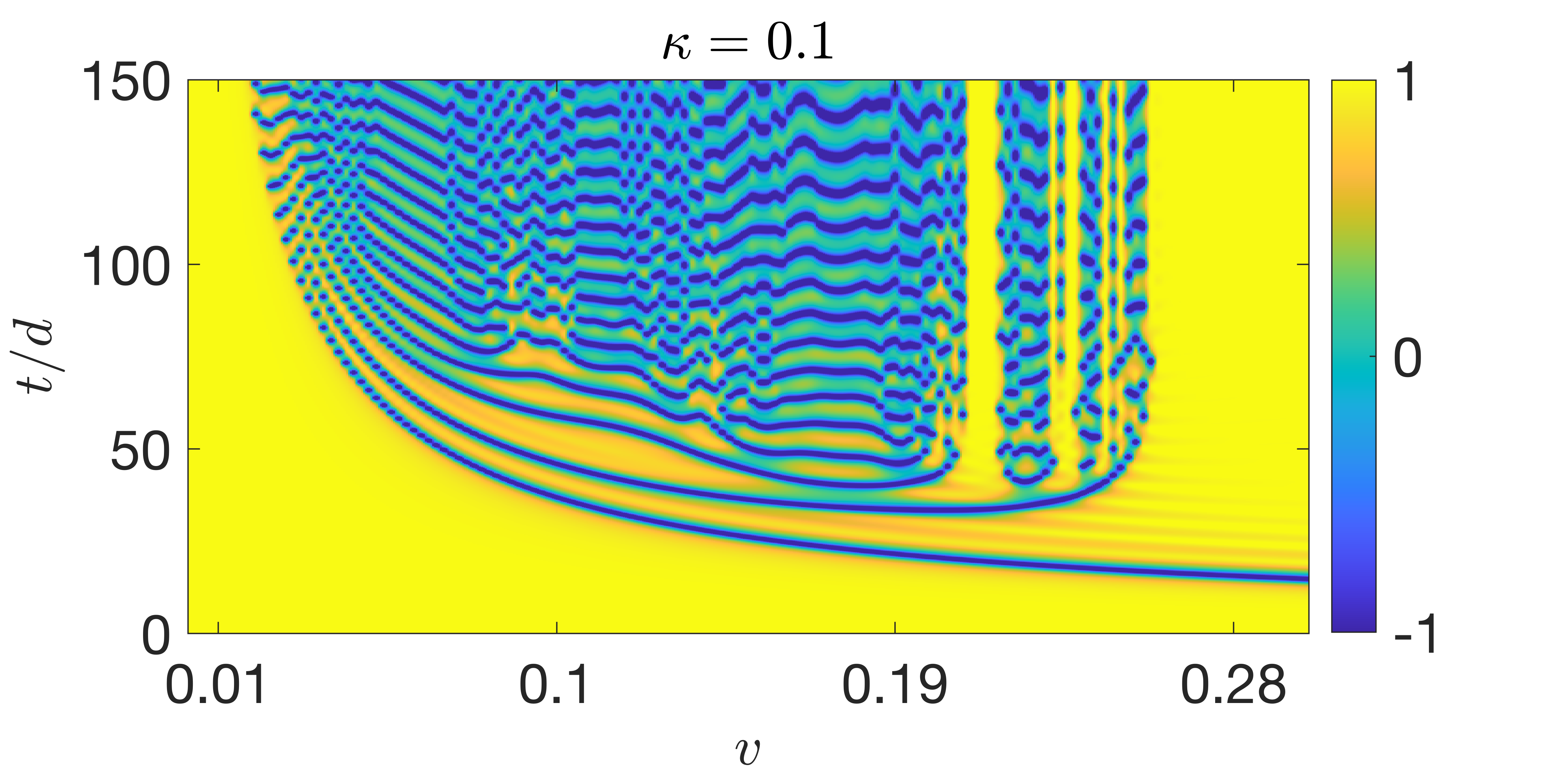}
    \includegraphics[width=\linewidth]{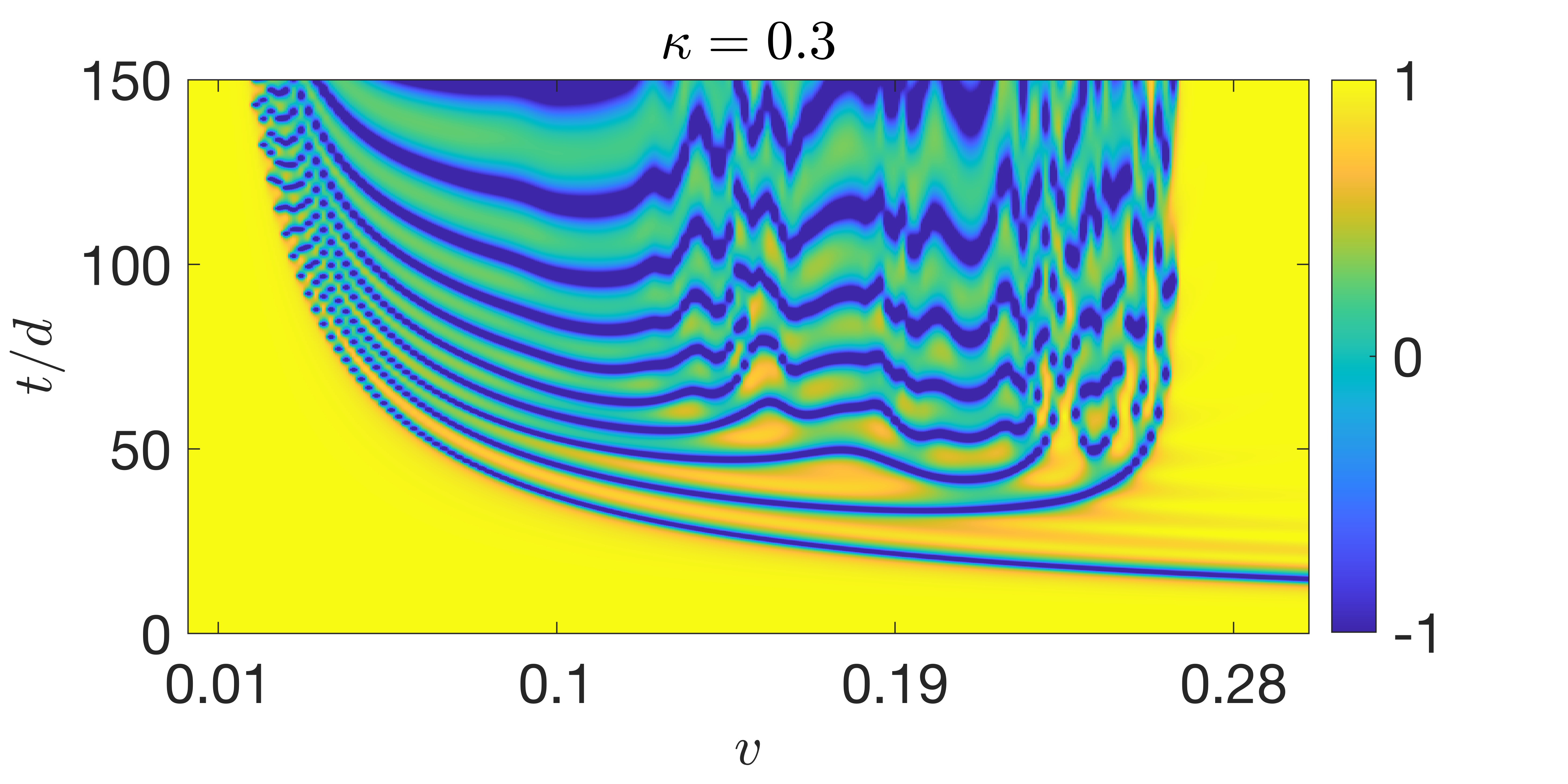}
    \includegraphics[width=\linewidth]{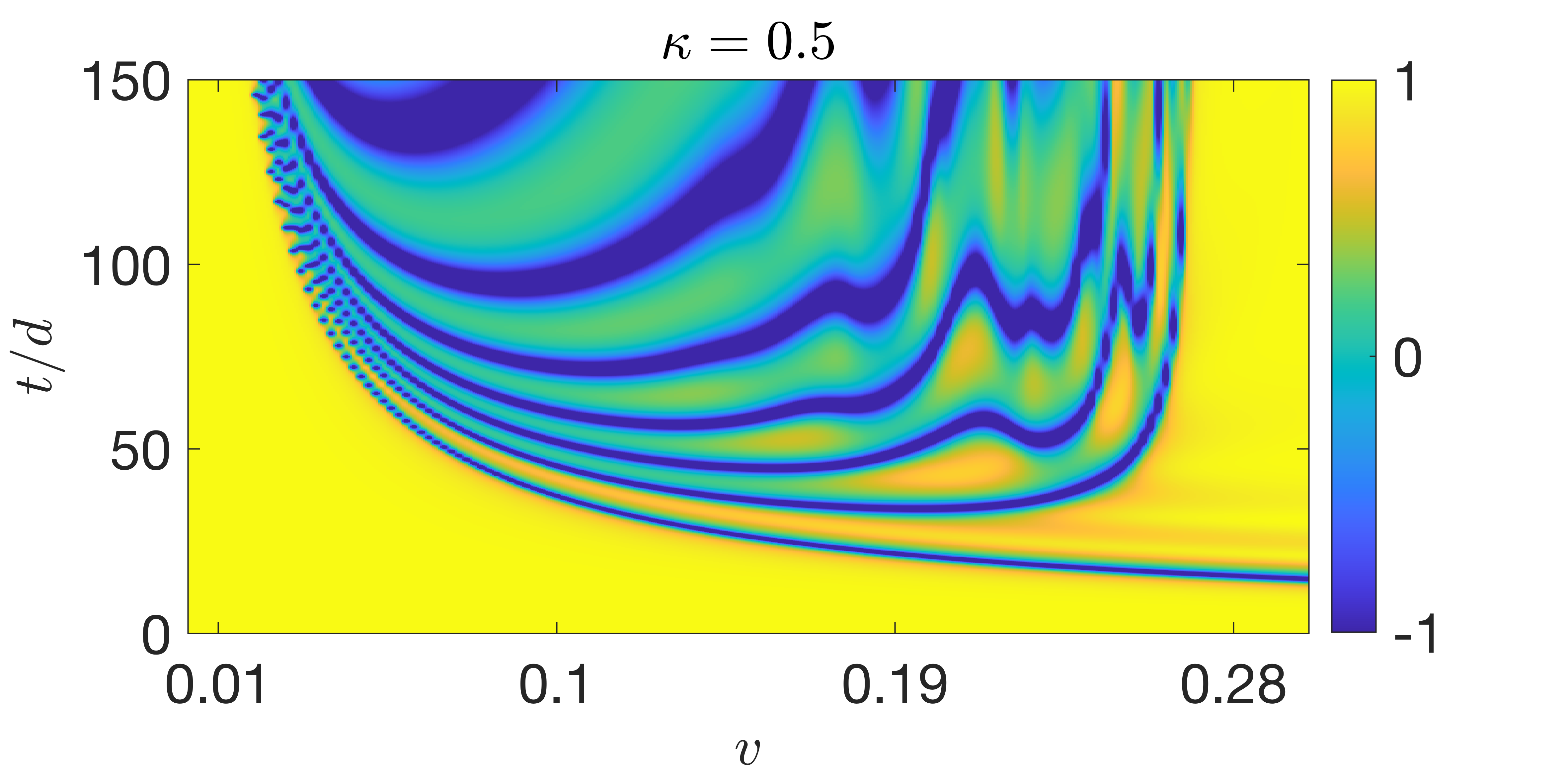}
    \caption{$\phi(t,z=0)$ for $\kappa=0.1, 0.3, 0.5$ and $v\in[0.002,0.3]$ with step size $\Delta v=0.002$.
As $\kappa$ increases, the bounce windows become narrower and shift toward higher initial velocities, and they are difficult to resolve in the strong-coupling regime. The critical velocity also increases with $\kappa$, although only mildly.}
 \label{scanv}
\end{figure}

Using the numerical methods described above, we scan the initial velocity from $v=0.002$ to $0.3$ with step size $\Delta v=0.002$ for several values of the gravitational coupling, $\kappa=0.1, 0.3$, and $0.5$. Figure~\ref{scanv} shows the time evolution of $\phi(t,z=0)$ as the initial velocity and gravitational coupling are varied. The figure indicates that increasing $\kappa$ tends to narrow the bounce windows and shift them toward larger initial velocities. Here, a bounce window refers to an interval of initial velocities for which the kink-antikink pair undergoes multiple collisions before eventually escaping. In the stronger-coupling regime, these windows become difficult to resolve clearly.

For comparison, in flat spacetime the first five two-bounce windows are approximately
$v\in[0.1926,0.2034]$, $[0.2241,0.2288]$, $[0.2372,0.2396]$, $[0.2440,0.2454]$, and $[0.2481,0.2490]$, see Ref.~\cite{Campbell1983}. With weak gravitational coupling, $\kappa=0.1$, representative two-bounce velocities are found near $v=0.215$, $0.235$, $0.245$, $0.251$, and $0.2535$, respectively. These representative velocities lie above the corresponding flat-spacetime intervals, indicating that gravity shifts the resonance windows toward higher initial velocities.

\subsection{Long-lived resonant quasi-bound states}
In flat spacetime, the $\phi^4$ kink possesses a translational zero mode and a localized shape mode. Resonant energy exchange between the translational motion of a kink-antikink pair and the shape mode provides the standard explanation for multi-bounce windows~\cite{Campbell1983}. This mechanism cannot be carried over directly to the present self-gravitating system. For $\kappa>0$, the linear perturbation equation about an arbitrary static solution is~\cite{Zhong2021}
\begin{equation}
\ddot{\delta\phi}-\delta\phi^{\prime\prime}
 +V_{\mathrm{eff}}(z)\delta\phi=0,
\end{equation}
where
\begin{equation}
V_{\mathrm{eff}}(z)=\frac{f^{\prime\prime}}{f},
\qquad f=\frac{\phi^{\prime}}{2A^{\prime}}.
\end{equation}
With the mode ansatz $\delta\phi(t,z)\sim\ee^{\ii\omega t}\psi(z)$, this becomes the Schr\"odinger-like equation
\begin{equation}
\hat{H}\psi\equiv-\psi^{\prime\prime}+V_{\mathrm{eff}}(z)\psi=\omega^2\psi.
\end{equation}

The stability operator can be factorized as
\begin{equation}
\hat{H}=\hat{\mathcal{A}}\hat{\mathcal{A}}^{\dagger},
\quad
\hat{\mathcal{A}}=\frac{\mathrm{d}}{\mathrm{d}z}+\frac{f'}{f},
\quad
\hat{\mathcal{A}}^{\dagger}=-\frac{\mathrm{d}}{\mathrm{d}z}+\frac{f'}{f},
\end{equation}
which implies that the spectrum of the self-adjoint stability operator is non-negative~\cite{Zhong2021}.

\begin{figure}[t]
    \centering
    \includegraphics[width=1\linewidth]{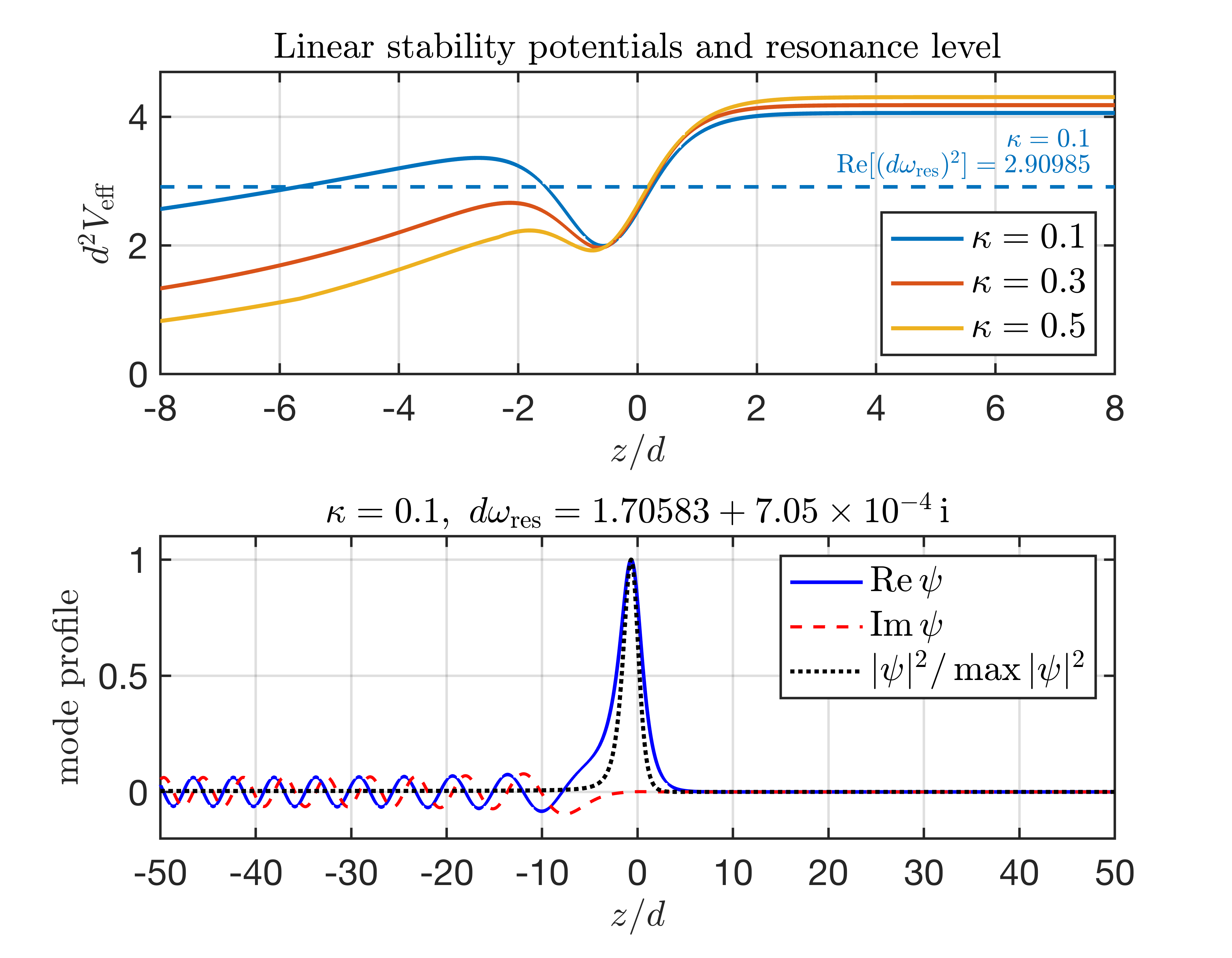}
    \caption{Linear stability potentials and the long-lived open-channel resonance for a kink. The upper panel shows $d^2V_{\mathrm{eff}}$ for $\kappa=0.1$, $0.3$, and $0.5$, together with the resonance level $\operatorname{Re}[(d\omega_{\mathrm{res}})^2]$ found at $\kappa=0.1$. The lower panel shows the real and imaginary parts of the corresponding complex mode profile and the rescaled quantity $|\psi|^2/\max_z|\psi(z)|^2$. The result is expressed in terms of $z/d$ and therefore applies to arbitrary $d$. The mode profile is rescaled for visualization.}
    \label{stabilityPotential}
\end{figure}

As can be seen in the upper panel of Fig.~\ref{stabilityPotential}, for a static kink $V_{\mathrm{eff}}\to0$ as $z\to-\infty$, so the continuous spectrum begins at $\omega^2=0$. The stability operator therefore has no eigenvalue below the continuum threshold, and its formal zero-energy solution $\psi_0\propto f$ is non-normalizable~\cite{Zhong2021}. Thus, unlike the flat-space kink, the gravitating kink has no normalizable discrete mode. Nevertheless, the weak-gravity potential can temporarily confine a perturbation near the kink while allowing it to leak into the left asymptotic region, thereby producing an open-channel resonance. 

We determine the complex resonance frequency by imposing boundary conditions for quasi-bound states: an outgoing wave in the open channel on the left and an evanescent tail in the closed channel on the right. Since $V_{\mathrm{eff}}(-\infty)=0$, the left asymptotic wave number satisfies $k_-^2=\omega^2$. Choosing the outgoing branch $k_-=\omega$ with $\operatorname{Re}\omega>0$, we obtain
\begin{equation}
\psi(z)\sim
\begin{cases}
\exp(\ii \omega z), & z\to-\infty,\\
\exp(-q_+z), & z\to+\infty,
\end{cases}
\end{equation}
where
\begin{equation}
q_+^2=V_{\mathrm{eff}}(+\infty)-\omega^2,\quad \operatorname{Re}q_+>0.
\end{equation}

We solve this boundary-value problem by integrating the equation from the two asymptotic regions toward a matching point and requiring the Wronskian of the two solutions to vanish there. 
For $\kappa=0.1$, the result is stable over the five cutoffs $z_{\max}/d=70$, $85$, $100$, $120$, and $140$, and yields
\begin{equation}
\label{eq:qbs}
d\omega_{\mathrm{res}}\simeq1.70583+7.05\times10^{-4}\ii.
\end{equation}
Note that the real part is close to the normal frequency $d\omega=\sqrt{3}\simeq1.732$ of the shape mode of $\phi^4$ kinks in the flat spacetime, suggesting that the quasi-bound states discovered here originate from the shape modes.
With the convention $\delta\phi\sim\ee^{\ii\omega t}\psi(z)$, the positive imaginary part corresponds to temporal decay. The associated quality factor is
\begin{equation}
Q=\frac{\operatorname{Re}\omega_{\mathrm{res}}}
{2|\operatorname{Im}\omega_{\mathrm{res}}|}
\simeq1.2\times10^3,
\end{equation}
indicating a very long-lived excitation of the self-gravitating kink. The corresponding potential and mode profile are shown in Fig.~\ref{stabilityPotential}. The oscillatory tail on the left reflects the open channel, whereas the rapid decay on the right reflects the closed channel. 
For $\kappa=0.3$ and $0.5$, the lifetimes of the quasi-bound states are too short for them to be identified using the direct-integration method adopted here.

For the antikink, $\phi_{\bar{\mathrm{K}}}(z)=\phi_{\mathrm{K}}(-z)$ and $A_{\bar{\mathrm{K}}}(z)=A_{\mathrm{K}}(-z)$, so $f_{\bar{\mathrm{K}}}(z)=f_{\mathrm{K}}(-z)$ and hence $V_{\mathrm{eff}}^{(\bar{\mathrm{K}})}(z)=V_{\mathrm{eff}}^{(\mathrm{K})}(-z)$.
Thus, its resonance spectrum is identical to that of the kink, while the open and closed channels are interchanged; it therefore suffices to present the kink results.

For the initial data in Eqs.~\eqref{phiinit} and \eqref{Ainit}, the kink and antikink open channels point toward the left and right exterior AdS$_2$ regions, respectively.
Since escaping collisions correspond to reflection rather than transmission, the kink returns to the left and the antikink to the right; hence, whenever the individual resonances are well defined, the leakage remains outward toward both AdS$_2$ boundaries.

\begin{figure}[t]
    \centering
    \includegraphics[width=0.85\linewidth]{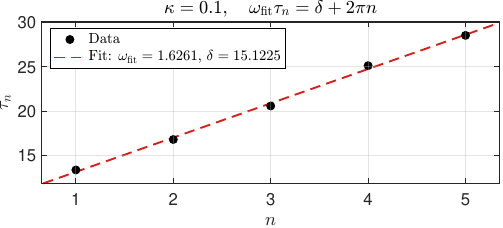}
    \caption{Proper time interval $\tau_n$ between two consecutive bounces as a function of the bounce window index $n$ for $\kappa=0.1$. The proper time is computed via $\tau_n = \int_{t_1}^{t_2} e^{A(t,z=0)}\,\mathrm{d}t$ between the two collision times. 
    Black dots show the extracted data for $n=1,\cdots,5$, and the red dashed line is the linear fit to $\omega_{\rm fit}\tau_n = \delta + 2\pi n$, yielding $\omega_{\rm fit} = 1.6261$ (we set $d=1$ here) and $\delta = 15.1225$.}
    \label{fig:2bouncefit}
\end{figure}

To test whether the long-lived resonant quasi-bound states can serve as an intermediate energy-storage mechanism in kink collisions, we estimate an effective oscillation frequency $\omega_{\rm fit}$ using the relation \cite{Campbell1983}
\begin{equation}
    \omega_{\rm fit} \tau_n = \delta +2\pi n
\end{equation}
with an offset phase $\delta$.
Note that $\tau_n =\int\ee^{A}\dd t$ is the proper time interval between two bounces in the $n$th two-bounce window, which reduces to the coordinate time interval in flat spacetime ($A\equiv0$).
As shown in Fig.~\ref{fig:2bouncefit}, such a linear fit yields $\omega_{\rm fit} = 1.6261$ (with $d=1$), close to the real part of \eqref{eq:qbs}.
However, the long-lived quasi-bound states, with a small imaginary part, could leak energy during collisions and may therefore contribute to the shift and narrowing of the windows and the increase of the critical velocity.
Similar phenomena were also reported in Ref.~\cite{Dorey2018}, where quasinormal modes that likewise originate from the shape mode act as dissipative energy-storage media.

In two dimensions, this leakage should not be interpreted as gravitational radiation in the usual higher-dimensional sense; it instead concerns radiative degrees of freedom in the coupled scalar-geometry system, which might be captured far away from the collision zone.
Extracting the corresponding asymptotic fluxes would require a more refined treatment, such as a hyperboloidal framework~\cite{Zenginoglu2024} or conformal field equations~\cite{Friedrich:1998xmu}, together with a suitable subtraction of the logarithmic asymptotic behavior of $A$.
Such an extraction is beyond the scope of the present work.

\subsection{Spacetime response to kink collisions}
\begin{figure*}[t]
    \centering
    \includegraphics[width=\linewidth]{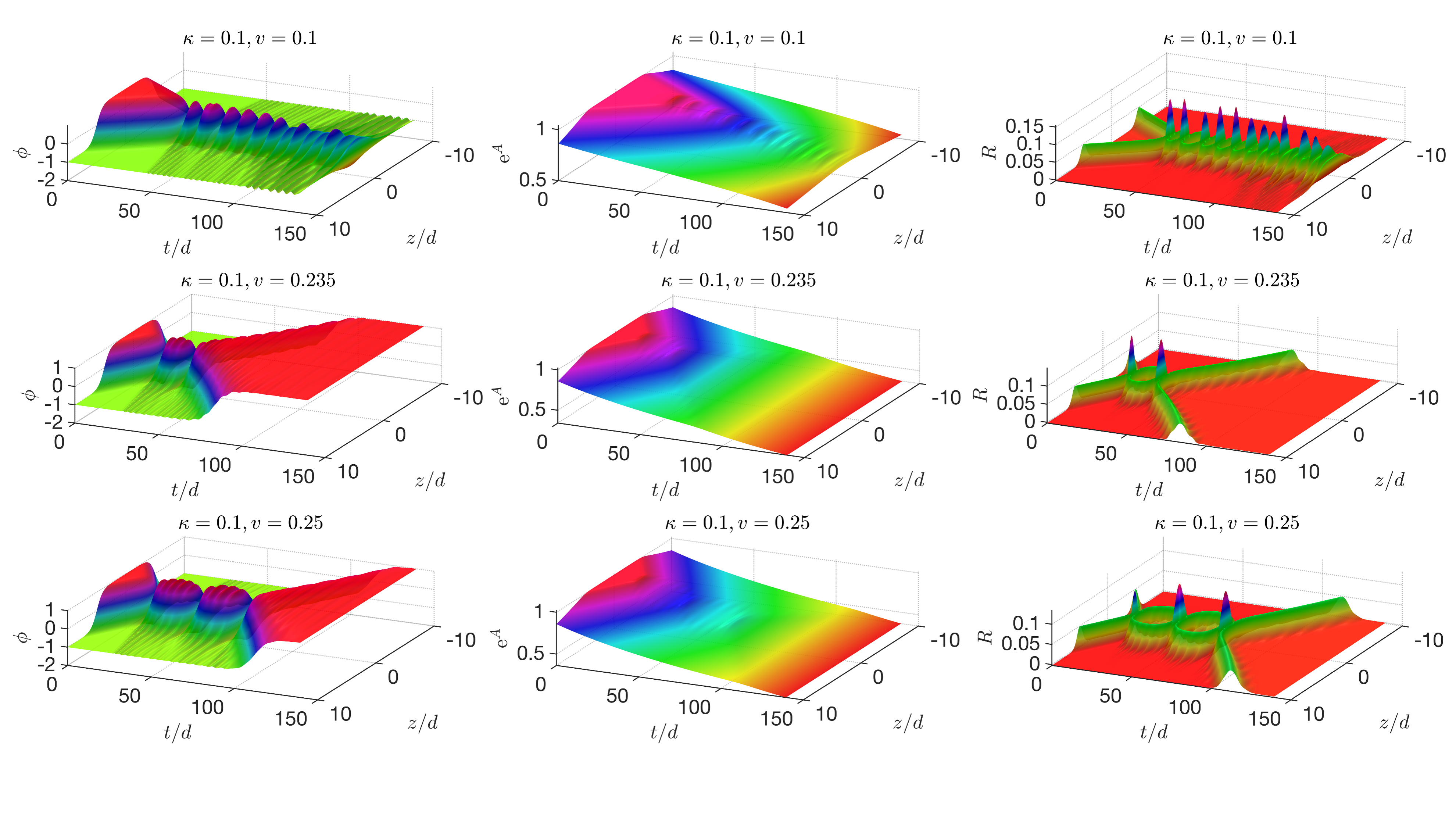}
\caption{Evolution of $\phi$ (left column), $\ee^{A}$ (middle column), and the Ricci scalar $R$ (right column) for $\kappa=0.1$ and representative initial velocities $v=0.1$, $0.235$, and $0.25$ from top to bottom.
The scalar-field profiles distinguish bion formation and multi-bounce escape, while $\ee^{A}$ and $R$ show the corresponding geometrical response.
The post-collision decrease of $\ee^{A}$ reflects a contraction of the local conformal spatial scale, and the finite curvature peaks indicate no sign of spacetime singularity formation in these examples.}
    \label{figK}
\end{figure*}

\begin{figure*}[t]
    \centering
    \includegraphics[width=1\linewidth]{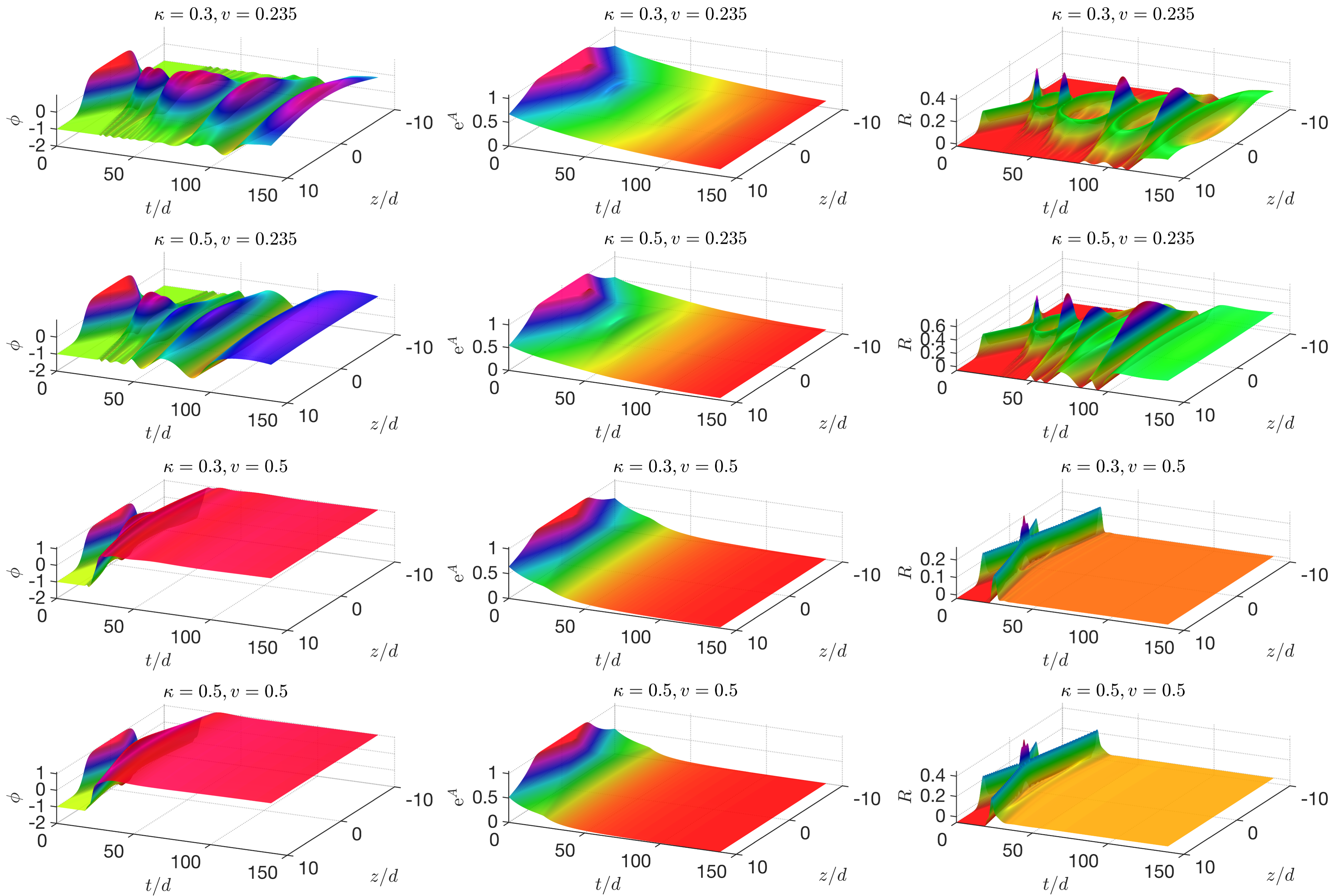}
\caption{Evolution of the scalar field $\phi$ (left column), the conformal factor $\ee^{A}$ (middle column), and the Ricci scalar $R$ (right column) for representative collisions with intermediate and strong gravitational coupling.
From top to bottom, the rows correspond to $(\kappa,v)=(0.3,0.235)$, $(0.5,0.235)$, $(0.3,0.5)$, and $(0.5,0.5)$.
The scalar-field profiles distinguish multi-collision dynamics at $v=0.235$ from single-bounce escape at $v=0.5$.
The conformal factor decreases after the collision, with a stronger decrease for larger $\kappa$, while the Ricci scalar develops transient but finite peaks during the collisions.
These features show that the geometrical backreaction becomes stronger as $\kappa$ increases, without evidence for spacetime singularity formation.}
    \label{fig:expA3D}
\end{figure*}

A notable feature of the evolution is that we do not observe any sign of spacetime singularity formation after the collision.
In our model, the Kretschmann scalar is simply related to the Ricci scalar by
\[
R_{abcd}R^{abcd}=R^2=(\kappa V)^2.
\]
Here the last equality follows from the field equations.
It is therefore sufficient to monitor the scalar curvature $R$ as a curvature diagnostic.
Representative evolutions of the scalar field, the conformal factor, and the Ricci scalar are shown in Figs.~\ref{figK} and~\ref{fig:expA3D}.
In all examples considered here, the Ricci scalar develops transient peaks during the collisions of the kink and antikink, but remains finite throughout the evolution.
This is in contrast to the case of five-dimensional thick branes~\cite{Takamizu2006,Takamizu2007,Omotani2011}, where curvature singularities may form soon after the collisions for sufficiently strong gravity.
The absence of singularity formation in our simulations may originate from differences in both the gravitational theory and the spacetime dimensionality, and warrants further investigation.

Figure~\ref{figK} illustrates the weak-gravity case $\kappa=0.1$.
The low-velocity collision, $v=0.1$, produces a long-lived bion state, while the cases $v=0.235$ and $v=0.25$ represent multi-bounce escape.
The repeated encounters visible in the scalar field are accompanied by recurrent curvature peaks, showing that the curvature response tracks the localization of scalar-field energy in the collision region.
At the same time, the warp factor $\ee^A$ decreases after the collision rather than returning to its initial profile, indicating that the geometry does not simply relax back to the original kink-antikink background.
Instead, the collision leaves a cumulative imprint on the conformal warp factor, reflecting the backreaction of the matter dynamics on the spacetime geometry.

In the conformal gauge, the proper distance associated with a fixed coordinate interval is $\dd l=\ee^A \dd z$.
Thus the post-collision decrease of $\ee^A$ corresponds to a contraction of the local proper spatial scale in this frame.
Equivalently, this behavior can be characterized by the expansion of the congruence of static observers,
\begin{equation}
\theta=\nabla_a u^a=\ee^{-A}\dot{A},
\end{equation}
where $u^a=(\ee^{-A},0)$.
The sign of $\theta$ shows whether the local conformal spatial scale is increasing or decreasing.
The contraction discussed here is therefore a local geometrical response in the chosen conformal frame, rather than a statement about a cosmological scale factor.

The dependence on $\kappa$ and $v$ is further illustrated in Fig.~\ref{fig:expA3D}.
For fixed velocity $v=0.235$, increasing $\kappa$ strengthens the geometrical response: the post-collision decrease of $\ee^A$ becomes more pronounced, and the curvature peaks become larger.
For the higher velocity $v=0.5$, the kink and antikink escape after a single collision, and the curvature response is correspondingly more localized in time.
Nevertheless, the same qualitative pattern persists: the collision produces a transient high-curvature region and a lasting decrease of the local conformal scale, but no curvature divergence is observed.
These results show that the self-gravity of the kink-antikink system affects not only the scattering outcome but also the post-collision spacetime geometry.

\section{Conclusions}\label{conclusion}
We have studied kink-antikink collisions in the self-gravitating $\phi^4$ model coupled to two-dimensional MMSS dilaton gravity. 
In this model, static self-gravitating kinks can be viewed as two-dimensional analogues of thick branes, providing a simple setting in which nonlinear scalar dynamics and gravitational backreaction can be followed simultaneously.

Our simulations show that gravity qualitatively modifies the scattering structure of the kink-antikink system. 
As the gravitational coupling $\kappa$ increases, the resonance windows shift toward higher initial velocities and become progressively narrower, while the critical velocity for escape increases mildly. 
Although the usual flat-spacetime vibrational-mode mechanism is not directly applicable to self-gravitating kinks, our linear perturbation analysis suggests that long-lived quasi-bound states provide a plausible energy-storage channel in the weak-gravity regime.

The collisions also produce a clear geometrical response. 
The conformal factor $\ee^A$ decreases after the collision, corresponding in the conformal gauge to a contraction of the local proper spatial scale, and this effect becomes stronger for larger $\kappa$. 
Meanwhile, the Ricci scalar develops transient peaks during the encounters but remains finite in all simulations considered here. 
Thus, in contrast to higher-dimensional thick-brane collisions, we find no evidence for spacetime singularity formation in the present two-dimensional model.

These results suggest several directions for future work, including a quantitative extraction of scalar and dilaton-metric perturbative fluxes, a more detailed study of resonant-mode excitation in nonlinear collisions, the construction of smooth, constraint-satisfying initial data for an antikink-kink configuration with asymptotically flat exterior regions, possible constraint-control methods such as that of Ref.~\cite{Vachaspati2016}, and extensions to other self-gravitating kink models or two-dimensional gravity theories.

The long-lived open-channel resonance identified here also motivates a quantum treatment of self-gravitating kink dynamics.
In flat-spacetime $\phi^4$ theory, quantum studies have investigated the unbinding, excitation, and decay of the kink shape mode, together with elastic and inelastic kink-meson scattering~\cite{Evslin2021,Evslin2022,Evslin2023,Evslin2024}.
A recent analysis further identified a Breit-Wigner resonance associated with a twice-excited shape mode in elastic kink-meson scattering~\cite{Bayarsaikhan2026}.
Extending such methods to the present gravitating background could clarify whether the classical quasi-bound excitation persists as a well-defined quantum resonance, how it interacts with continuum quantum modes, and how gravitational backreaction modifies its lifetime and scattering signatures.

\section*{Acknowledgments}
We thank Kei-ichi Maeda and Yong-Qiang Wang for helpful discussions.
Z.-T. H. acknowledges the Xiaomi MiMo Orbit 100T Token Grant for Builders for providing 700 million computational credits, which supported the implementation and testing of the numerical methods used in this work.
Part of this work was completed during Z.-T. H.'s undergraduate years and was presented as his bachelor's thesis in Chinese.
This work was supported by the National Natural Science Foundation of China (Grant number 12175169).

\bibliographystyle{elsarticle-num}

\begin{thebibliography}{100}
\expandafter\ifx\csname url\endcsname\relax
  \def\url#1{\texttt{#1}}\fi
\expandafter\ifx\csname urlprefix\endcsname\relax\def\urlprefix{URL }\fi
\expandafter\ifx\csname href\endcsname\relax
  \def\href#1#2{#2} \def\path#1{#1}\fi

\bibitem{Vachaspati2007}
T.~Vachaspati, {Kinks and domain walls: an introduction to classical and
  quantum solitons}, Cambridge University Press, 2006.
\newblock \href {https://doi.org/10.1017/9781009290456}
  {\path{doi:10.1017/9781009290456}}.

\bibitem{Campo2014}
A.~del Campo, W.~H. Zurek, {Universality of phase transition dynamics:
  topological defects from symmetry breaking}, Int. J. Mod. Phys. A 29~(8)
  (2014) 1430018.
\newblock \href {http://arxiv.org/abs/1310.1600} {\path{arXiv:1310.1600}},
  \href {https://doi.org/10.1142/S0217751X1430018X}
  {\path{doi:10.1142/S0217751X1430018X}}.

\bibitem{Kibble1976}
T.~W.~B. Kibble, {Topology of cosmic domains and strings}, J. Phys. A 9 (1976)
  1387--1398.
\newblock \href {https://doi.org/10.1088/0305-4470/9/8/029}
  {\path{doi:10.1088/0305-4470/9/8/029}}.

\bibitem{Kibble1980}
T.~W.~B. Kibble, {Some implications of a cosmological phase transition}, Phys.
  Rept. 67 (1980) 183.
\newblock \href {https://doi.org/10.1016/0370-1573(80)90091-5}
  {\path{doi:10.1016/0370-1573(80)90091-5}}.

\bibitem{Vilenkin1981}
A.~Vilenkin, {Cosmic strings}, Phys. Rev. D 24 (1981) 2082--2089.
\newblock \href {https://doi.org/10.1103/PhysRevD.24.2082}
  {\path{doi:10.1103/PhysRevD.24.2082}}.

%
%

\bibitem{Randall1999}
L.~Randall, R.~Sundrum, {A large mass hierarchy from a small extra dimension},
  Phys. Rev. Lett. 83 (1999) 3370--3373.
\newblock \href {http://arxiv.org/abs/hep-ph/9905221}
  {\path{arXiv:hep-ph/9905221}}, \href
  {https://doi.org/10.1103/PhysRevLett.83.3370}
  {\path{doi:10.1103/PhysRevLett.83.3370}}.

\bibitem{Randall1999a}
L.~Randall, R.~Sundrum, {An alternative to compactification}, Phys. Rev. Lett.
  83 (1999) 4690--4693.
\newblock \href {http://arxiv.org/abs/hep-th/9906064}
  {\path{arXiv:hep-th/9906064}}, \href
  {https://doi.org/10.1103/PhysRevLett.83.4690}
  {\path{doi:10.1103/PhysRevLett.83.4690}}.

\bibitem{Skenderis:1999mm}
K.~Skenderis, P.~K. Townsend, {Gravitational stability and renormalization
  group flow}, Phys. Lett. B 468 (1999) 46--51.
\newblock \href {http://arxiv.org/abs/hep-th/9909070}
  {\path{arXiv:hep-th/9909070}}, \href
  {https://doi.org/10.1016/S0370-2693(99)01212-5}
  {\path{doi:10.1016/S0370-2693(99)01212-5}}.

\bibitem{DeWolfe:1999cp}
O.~DeWolfe, D.~Z. Freedman, S.~S. Gubser, A.~Karch, {Modeling the
  fifth-dimension with scalars and gravity}, Phys. Rev. D 62 (2000) 046008.
\newblock \href {http://arxiv.org/abs/hep-th/9909134}
  {\path{arXiv:hep-th/9909134}}, \href
  {https://doi.org/10.1103/PhysRevD.62.046008}
  {\path{doi:10.1103/PhysRevD.62.046008}}.

\bibitem{Gremm:1999pj}
M.~Gremm, {Four-dimensional gravity on a thick domain wall}, Phys. Lett. B 478
  (2000) 434--438.
\newblock \href {http://arxiv.org/abs/hep-th/9912060}
  {\path{arXiv:hep-th/9912060}}, \href
  {https://doi.org/10.1016/S0370-2693(00)00303-8}
  {\path{doi:10.1016/S0370-2693(00)00303-8}}.

\bibitem{Dzhunushaliev2010}
V.~Dzhunushaliev, V.~Folomeev, M.~Minamitsuji, {Thick brane solutions}, Rept.
  Prog. Phys. 73 (2010) 066901.
\newblock \href {http://arxiv.org/abs/0904.1775} {\path{arXiv:0904.1775}},
  \href {https://doi.org/10.1088/0034-4885/73/6/066901}
  {\path{doi:10.1088/0034-4885/73/6/066901}}.

\bibitem{Liu2018}
Y.-X. Liu, {Introduction to extra dimensions and thick braneworlds}, World
  Scientific, 2018, Ch.~8, pp. 211--275.
\newblock \href {http://arxiv.org/abs/1707.08541} {\path{arXiv:1707.08541}},
  \href {https://doi.org/10.1142/9789813237278_0008}
  {\path{doi:10.1142/9789813237278_0008}}.

\bibitem{Sugiyama1979}
T.~Sugiyama, {Kink-antikink collisions in the two-dimensional $\varphi^4$ model},
  Prog. Theor. Phys. 61 (1979) 1550--1563.
\newblock \href {https://doi.org/10.1143/PTP.61.1550}
  {\path{doi:10.1143/PTP.61.1550}}.

\bibitem{Moshir1981}
M.~Moshir, {Soliton-antisoliton scattering and capture in $\lambda \varphi^4$
  theory}, Nucl. Phys. B 185 (1981) 318--332.
\newblock \href {https://doi.org/10.1016/0550-3213(81)90320-5}
  {\path{doi:10.1016/0550-3213(81)90320-5}}.

\bibitem{Campbell1983}
D.~K. Campbell, J.~F. Schonfeld, C.~A. Wingate, {Resonance structure in
  kink-antikink interactions in $\varphi^4$ theory}, Physica D 9 (1983) 1.
\newblock \href {https://doi.org/10.1016/0167-2789(83)90289-0}
  {\path{doi:10.1016/0167-2789(83)90289-0}}.

\bibitem{Anninos1991}
P.~Anninos, S.~Oliveira, R.~A. Matzner, {Fractal structure in the scalar
  $\lambda(\varphi^2-1)^2$ theory}, Phys. Rev. D 44 (1991) 1147--1160.
\newblock \href {https://doi.org/10.1103/PhysRevD.44.1147}
  {\path{doi:10.1103/PhysRevD.44.1147}}.

\bibitem{Goodman2007}
R.~H. Goodman, R.~Haberman,
  \href{https://link.aps.org/doi/10.1103/PhysRevLett.98.104103}{Chaotic
  scattering and the $n$-bounce resonance in solitary-wave interactions}, Phys.
  Rev. Lett. 98 (2007) 104103.
\newblock \href {https://doi.org/10.1103/PhysRevLett.98.104103}
  {\path{doi:10.1103/PhysRevLett.98.104103}}.
\newline\urlprefix\url{https://link.aps.org/doi/10.1103/PhysRevLett.98.104103}

\bibitem{Dorey2017}
P.~Dorey, A.~Halavanau, J.~Mercer, T.~Roma{\'n}czukiewicz, Y.~Shnir, {Boundary
  scattering in the $\phi^4$ model}, JHEP 05 (2017) 107.
\newblock \href {http://arxiv.org/abs/1508.02329} {\path{arXiv:1508.02329}},
  \href {https://doi.org/10.1007/JHEP05(2017)107}
  {\path{doi:10.1007/JHEP05(2017)107}}.

\bibitem{Campbell1986}
D.~K. Campbell, M.~Peyrard, P.~Sodano, {Kink-antikink interactions in the
  double sine-Gordon equation}, Physica D 19~(2) (1986) 165--205.
\newblock \href {https://doi.org/10.1016/0167-2789(86)90019-9}
  {\path{doi:10.1016/0167-2789(86)90019-9}}.

\bibitem{Gani1999}
V.~A. Gani, A.~E. Kudryavtsev, {Kink-antikink interactions in the double
  sine-Gordon equation and the problem of resonance frequencies}, Phys. Rev. E
  60 (1999) 3305--3309.
\newblock \href {http://arxiv.org/abs/cond-mat/9809015}
  {\path{arXiv:cond-mat/9809015}}, \href
  {https://doi.org/10.1103/PhysRevE.60.3305}
  {\path{doi:10.1103/PhysRevE.60.3305}}.

\bibitem{Gani2022}
V.~A. Gani, A.~Gorina, I.~Perapechka, Y.~Shnir, {Remarks on sine-Gordon
  kink-fermion system: localized modes and scattering}, Eur. Phys. J. C 82~(8)
  (2022) 757.
\newblock \href {http://arxiv.org/abs/2205.13437} {\path{arXiv:2205.13437}},
  \href {https://doi.org/10.1140/epjc/s10052-022-10707-0}
  {\path{doi:10.1140/epjc/s10052-022-10707-0}}.


\bibitem{Dorey2021}
P.~Dorey, A.~Gorina, I.~Perapechka, T.~Roma{\'n}czukiewicz, Y.~Shnir,
  {Resonance structures in kink-antikink collisions in a deformed sine-Gordon
  model}, JHEP 09 (2021) 145.
\newblock \href {http://arxiv.org/abs/2106.09560} {\path{arXiv:2106.09560}},
  \href {https://doi.org/10.1007/JHEP09(2021)145}
  {\path{doi:10.1007/JHEP09(2021)145}}.

\bibitem{CarreteroGonzalez2022}
R.~Carretero-Gonzalez, L.~A. Cisneros-Ake, R.~Decker, G.~N. Koutsokostas, D.~J.
  Frantzeskakis, P.~G. Kevrekidis, D.~J. Ratliff, {Kink{\textendash}antikink
  stripe interactions in the two-dimensional sine{\textendash}Gordon equation},
  Commun. Nonlinear Sci. Numer. Simul. 109 (2022) 106123.
\newblock \href {http://arxiv.org/abs/2108.03121} {\path{arXiv:2108.03121}},
  \href {https://doi.org/10.1016/j.cnsns.2021.106123}
  {\path{doi:10.1016/j.cnsns.2021.106123}}.

\bibitem{Hora2025}
E.~da~Hora, L.~Pereira, C.~dos Santos, F.~C. Simas, {Geometrically constrained
  sine-Gordon field: BPS solitons and their collisions}, Commun. Nonlinear Sci.
  Numer. Simul. 151 (2025) 109070.
\newblock \href {http://arxiv.org/abs/2409.09767} {\path{arXiv:2409.09767}},
  \href {https://doi.org/10.1016/j.cnsns.2025.109070}
  {\path{doi:10.1016/j.cnsns.2025.109070}}.

\bibitem{Hoseinmardy2010}
S.~Hoseinmardy, N.~Riazi, {Inelastic collision of kinks and antikinks in the
  $\phi^6$ system}, Int. J. Mod. Phys. A 25 (2010) 3261--3270.
\newblock \href {https://doi.org/10.1142/S0217751X10049712}
  {\path{doi:10.1142/S0217751X10049712}}.

\bibitem{Belendryasova2017}
E.~Belendryasova, V.~A. Gani, {Resonance phenomena in the $\varphi^8$ kink
  scattering}, J. Phys. Conf. Ser. 934~(1) (2017) 012059.
\newblock \href {http://arxiv.org/abs/1712.02846} {\path{arXiv:1712.02846}},
  \href {https://doi.org/10.1088/1742-6596/934/1/012059}
  {\path{doi:10.1088/1742-6596/934/1/012059}}.

\bibitem{Gani2021}
V.~A. Gani, A.~M. Marjaneh, K.~Javidan, {Exotic final states in the $\varphi^8$
  multi-kink collisions}, Eur. Phys. J. C 81~(12) (2021) 1124.
\newblock \href {http://arxiv.org/abs/2106.06399} {\path{arXiv:2106.06399}},
  \href {https://doi.org/10.1140/epjc/s10052-021-09935-7}
  {\path{doi:10.1140/epjc/s10052-021-09935-7}}.

\bibitem{Khare2022}
A.~Khare, A.~Saxena, {Kink solutions with power law tails}, Front. Phys. 10
  (2022) 992915.
\newblock \href {http://arxiv.org/abs/2207.10876} {\path{arXiv:2207.10876}},
  \href {https://doi.org/10.3389/fphy.2022.992915}
  {\path{doi:10.3389/fphy.2022.992915}}.

\bibitem{MoradiMarjaneh2022}
A.~Moradi~Marjaneh, F.~C. Simas, D.~Bazeia, {Collisions of kinks in deformed
  $\varphi^4$ and $\varphi^6$ models}, Chaos Solitons Fractals 164 (2022)
  112723.
\newblock \href {http://arxiv.org/abs/2207.00835} {\path{arXiv:2207.00835}},
  \href {https://doi.org/10.1016/j.chaos.2022.112723}
  {\path{doi:10.1016/j.chaos.2022.112723}}.

\bibitem{Bazeia2023}
D.~Bazeia, J.~G.~F. Campos, A.~Mohammadi, {Kink-antikink collisions in the
  $\phi^8$ model: short-range to long-range journey}, JHEP 05 (2023) 116.
\newblock \href {http://arxiv.org/abs/2303.12482} {\path{arXiv:2303.12482}},
  \href {https://doi.org/10.1007/JHEP05(2023)116}
  {\path{doi:10.1007/JHEP05(2023)116}}.

\bibitem{Belendryasova2021}
E.~Belendryasova, V.~A. Gani, K.~G. Zloshchastiev, {Kink solutions in
  logarithmic scalar field theory: excitation spectra, scattering, and decay of
  bions}, Phys. Lett. B 823 (2021) 136776.
\newblock \href {http://arxiv.org/abs/2111.09096} {\path{arXiv:2111.09096}},
  \href {https://doi.org/10.1016/j.physletb.2021.136776}
  {\path{doi:10.1016/j.physletb.2021.136776}}.

\bibitem{Campos2021}
J.~G.~F. Campos, A.~Mohammadi, {Wobbling double sine-Gordon kinks}, JHEP 09
  (2021) 067.
\newblock \href {http://arxiv.org/abs/2103.04908} {\path{arXiv:2103.04908}},
  \href {https://doi.org/10.1007/JHEP09(2021)067}
  {\path{doi:10.1007/JHEP09(2021)067}}.

\bibitem{AlonsoIzquierdo2021}
A.~Alonso~Izquierdo, L.~M. Nieto, J.~Queiroga-Nunes, {Scattering between
  wobbling kinks}, Phys. Rev. D 103~(4) (2021) 045003.
\newblock \href {http://arxiv.org/abs/2007.15517} {\path{arXiv:2007.15517}},
  \href {https://doi.org/10.1103/PhysRevD.103.045003}
  {\path{doi:10.1103/PhysRevD.103.045003}}.

\bibitem{AlonsoIzquierdo2022}
A.~Alonso-Izquierdo, L.~M. Nieto, J.~Queiroga-Nunes, {Asymmetric scattering
  between kinks and wobblers}, Commun. Nonlinear Sci. Numer. Simul. 107 (2022)
  106183.
\newblock \href {http://arxiv.org/abs/2109.13904} {\path{arXiv:2109.13904}},
  \href {https://doi.org/10.1016/j.cnsns.2021.106183}
  {\path{doi:10.1016/j.cnsns.2021.106183}}.

\bibitem{Kivshar1991}
Y.~S. Kivshar, Z.~Fei, L.~V{\'a}zquez, {Resonant soliton-impurity
  interactions}, Phys. Rev. Lett. 67 (1991) 1177--1180.
\newblock \href {https://doi.org/10.1103/PhysRevLett.67.1177}
  {\path{doi:10.1103/PhysRevLett.67.1177}}.

\bibitem{Fei1992}
Z.~Fei, Y.~S. Kivshar, L.~V\'azquez,
  \href{https://link.aps.org/doi/10.1103/PhysRevA.45.6019}{Resonant
  kink-impurity interactions in the sine-gordon model}, Phys. Rev. A 45 (1992)
  6019--6030.
\newblock \href {https://doi.org/10.1103/PhysRevA.45.6019}
  {\path{doi:10.1103/PhysRevA.45.6019}}.
\newline\urlprefix\url{https://link.aps.org/doi/10.1103/PhysRevA.45.6019}

\bibitem{Zhou2017}
Y.~Zhou, B.~G.-g. Chen, N.~Upadhyaya, V.~Vitelli, {Kink-antikink asymmetry and
  impurity interactions in topological mechanical chains}, Phys. Rev. E 95~(2)
  (2017) 022202.
\newblock \href {http://arxiv.org/abs/1608.02127} {\path{arXiv:1608.02127}},
  \href {https://doi.org/10.1103/PhysRevE.95.022202}
  {\path{doi:10.1103/PhysRevE.95.022202}}.

\bibitem{Lizunova2021}
M.~Lizunova, J.~Kager, S.~de~Lange, J.~van Wezel, {Emergence of oscillons in
  kink-impurity interactions}, J. Phys. A 54 (2021) 315701.
\newblock \href {http://arxiv.org/abs/2012.07281} {\path{arXiv:2012.07281}},
  \href {https://doi.org/10.1088/1751-8121/ac0d36}
  {\path{doi:10.1088/1751-8121/ac0d36}}.

\bibitem{Zhong2020}
Y.~Zhong, X.-L. Du, Z.-C. Jiang, Y.-X. Liu, Y.-Q. Wang, {Collision of two kinks
  with inner structure}, JHEP 02 (2020) 153.
\newblock \href {http://arxiv.org/abs/1906.02920} {\path{arXiv:1906.02920}},
  \href {https://doi.org/10.1007/JHEP02(2020)153}
  {\path{doi:10.1007/JHEP02(2020)153}}.

\bibitem{Yan2020}
H.~Yan, Y.~Zhong, Y.-X. Liu, K.-i. Maeda, {Kink-antikink collision in a
  Lorentz-violating $\phi^4$ model}, Phys. Lett. B 807 (2020) 135542.
\newblock \href {http://arxiv.org/abs/2004.13329} {\path{arXiv:2004.13329}},
  \href {https://doi.org/10.1016/j.physletb.2020.135542}
  {\path{doi:10.1016/j.physletb.2020.135542}}.

\bibitem{Santos2025}
C.~E.~S. Santos, J.~G.~F. Campos, A.~Mohammadi, {On the localized and
  delocalized modes in kink-antikink interactions: a toy model}, JHEP 01 (2025)
  035.
\newblock \href {http://arxiv.org/abs/2408.00945} {\path{arXiv:2408.00945}},
  \href {https://doi.org/10.1007/JHEP01(2025)035}
  {\path{doi:10.1007/JHEP01(2025)035}}.

\bibitem{Dorey2011}
P.~Dorey, K.~Mersh, T.~Roma{\'n}czukiewicz, Y.~Shnir, {Kink-antikink collisions
  in the $\phi^6$ model}, Phys. Rev. Lett. 107 (2011) 091602.
\newblock \href {http://arxiv.org/abs/1101.5951} {\path{arXiv:1101.5951}},
  \href {https://doi.org/10.1103/PhysRevLett.107.091602}
  {\path{doi:10.1103/PhysRevLett.107.091602}}.

\bibitem{Gani2014}
V.~A. Gani, A.~E. Kudryavtsev, M.~A. Lizunova, {Kink interactions in the
  $(1+1)$-dimensional $\phi^6$ model}, Phys. Rev. D 89~(12) (2014) 125009.
\newblock \href {http://arxiv.org/abs/1402.5903} {\path{arXiv:1402.5903}},
  \href {https://doi.org/10.1103/PhysRevD.89.125009}
  {\path{doi:10.1103/PhysRevD.89.125009}}.

\bibitem{Yan2022}
H.~Yan, {Kink scattering in a Lorentz-violating $\phi^6$ model}, EPL 138 (2022)
  14001.
\newblock \href {http://arxiv.org/abs/2110.13381} {\path{arXiv:2110.13381}},
  \href {https://doi.org/10.1209/0295-5075/ac5b9b}
  {\path{doi:10.1209/0295-5075/ac5b9b}}.

\bibitem{Adam2022}
C.~Adam, P.~Dorey, A.~Garc{\'\i}a Mart{\'\i}n-Caro, M.~Huidobro, K.~Ole{\'s},
  T.~Roma{\'n}czukiewicz, Y.~Shnir, A.~Wereszczy{\'n}ski, {Multikink scattering
  in the $\phi^6$ model revisited}, Phys. Rev. D 106~(12) (2022) 125003.
\newblock \href {http://arxiv.org/abs/2209.08849} {\path{arXiv:2209.08849}},
  \href {https://doi.org/10.1103/PhysRevD.106.125003}
  {\path{doi:10.1103/PhysRevD.106.125003}}.

\bibitem{Dorey2018}
P.~Dorey, T.~Roma{\'n}czukiewicz, {Resonant kink-antikink scattering through
  quasinormal modes}, Phys. Lett. B 779 (2018) 117--123.
\newblock \href {http://arxiv.org/abs/1712.10235} {\path{arXiv:1712.10235}},
  \href {https://doi.org/10.1016/j.physletb.2018.02.003}
  {\path{doi:10.1016/j.physletb.2018.02.003}}.

\bibitem{Campos2020}
J.~G.~F. Campos, A.~Mohammadi, {Quasinormal modes in kink excitations and
  kink{\textendash}antikink interactions: a toy model}, Eur. Phys. J. C 80~(5)
  (2020) 352.
\newblock \href {http://arxiv.org/abs/1905.00835} {\path{arXiv:1905.00835}},
  \href {https://doi.org/10.1140/epjc/s10052-020-7856-3}
  {\path{doi:10.1140/epjc/s10052-020-7856-3}}.

\bibitem{Campos2022}
J.~G.~F. Campos, A.~Mohammadi, {Kink-antikink collision in the supersymmetric
  $\phi^4$ model}, JHEP 08 (2022) 180.
\newblock \href {http://arxiv.org/abs/2205.06869} {\path{arXiv:2205.06869}},
  \href {https://doi.org/10.1007/JHEP08(2022)180}
  {\path{doi:10.1007/JHEP08(2022)180}}.

\bibitem{Weigel2014}
H.~Weigel, {Kink-antikink scattering in $\varphi^4$ and $\phi^6$ models}, J.
  Phys. Conf. Ser. 482 (2014) 012045.
\newblock \href {http://arxiv.org/abs/1309.6607} {\path{arXiv:1309.6607}},
  \href {https://doi.org/10.1088/1742-6596/482/1/012045}
  {\path{doi:10.1088/1742-6596/482/1/012045}}.

\bibitem{Takyi2016}
I.~Takyi, H.~Weigel, {Collective coordinates in one-dimensional soliton models
  revisited}, Phys. Rev. D 94~(8) (2016) 085008.
\newblock \href {http://arxiv.org/abs/1609.06833} {\path{arXiv:1609.06833}},
  \href {https://doi.org/10.1103/PhysRevD.94.085008}
  {\path{doi:10.1103/PhysRevD.94.085008}}.

\bibitem{Weigel2018}
H.~Weigel, {Collective coordinate methods and their applicability to
  $\varphi^4$ models}, Springer International Publishing, Cham, 2019, Ch.~3,
  pp. 51--74.
\newblock \href {http://arxiv.org/abs/1809.03772} {\path{arXiv:1809.03772}},
  \href {https://doi.org/10.1007/978-3-030-11839-6\_3}
  {\path{doi:10.1007/978-3-030-11839-6\_3}}.

\bibitem{Manton2021}
N.~S. Manton, K.~Ole{\'s}, T.~Roma{\'n}czukiewicz, A.~Wereszczy{\'n}ski,
  {Collective coordinate model of kink-antikink collisions in $\phi^4$ theory},
  Phys. Rev. Lett. 127~(7) (2021) 071601.
\newblock \href {http://arxiv.org/abs/2106.05153} {\path{arXiv:2106.05153}},
  \href {https://doi.org/10.1103/PhysRevLett.127.071601}
  {\path{doi:10.1103/PhysRevLett.127.071601}}.

\bibitem{Manton2021a}
N.~S. Manton, K.~Ole{\'s}, T.~Roma{\'n}czukiewicz, A.~Wereszczy{\'n}ski, {Kink
  moduli spaces: collective coordinates reconsidered}, Phys. Rev. D 103~(2)
  (2021) 025024.
\newblock \href {http://arxiv.org/abs/2008.01026} {\path{arXiv:2008.01026}},
  \href {https://doi.org/10.1103/PhysRevD.103.025024}
  {\path{doi:10.1103/PhysRevD.103.025024}}.

\bibitem{Adam2022a}
C.~Adam, N.~S. Manton, K.~Ole{\'s}, T.~Roma{\'n}czukiewicz,
  A.~Wereszczy{\'n}ski, {Relativistic moduli space for kink collisions}, Phys.
  Rev. D 105~(6) (2022) 065012.
\newblock \href {http://arxiv.org/abs/2111.06790} {\path{arXiv:2111.06790}},
  \href {https://doi.org/10.1103/PhysRevD.105.065012}
  {\path{doi:10.1103/PhysRevD.105.065012}}.

\bibitem{Pereira2021}
C.~F.~S. Pereira, G.~Luchini, T.~Tassis, C.~P. Constantinidis, {Some novel
  considerations about the collective coordinates approximation for the
  scattering of $\phi^4$ kinks}, J. Phys. A: Math. Theor. 54~(7) (2021)
  075701.
\newblock \href {http://arxiv.org/abs/2004.00571} {\path{arXiv:2004.00571}},
  \href {https://doi.org/10.1088/1751-8121/abd815}
  {\path{doi:10.1088/1751-8121/abd815}}.

\bibitem{Pereira2023}
C.~F.~S. Pereira, E.~dos Santos Costa~Filho, T.~Tassis, {Collective
  coordinates for the hybrid model}, Int. J. Mod. Phys. A 38~(1) (2023)
  2350006.
\newblock \href {http://arxiv.org/abs/2110.05658} {\path{arXiv:2110.05658}},
  \href {https://doi.org/10.1142/S0217751X23500069}
  {\path{doi:10.1142/S0217751X23500069}}.

\bibitem{RoyH.2006}
R.~H. Goodman, R.~Haberman, {Kink-antikink collisions in the $\phi^4$ equation:
  the $n$-bounce resonance and the separatrix map}, SIAM J. Appl. Dynam. Syst.
  4~(4) (2005) 1195--1228.
\newblock \href {https://doi.org/10.1137/050632981}
  {\path{doi:10.1137/050632981}}.

\bibitem{Hawking1982}
S.~W. Hawking, I.~G. Moss, J.~M. Stewart,
  \href{https://link.aps.org/doi/10.1103/PhysRevD.26.2681}{Bubble collisions in
  the very early universe}, Phys. Rev. D 26 (1982) 2681--2693.
\newblock \href {https://doi.org/10.1103/PhysRevD.26.2681}
  {\path{doi:10.1103/PhysRevD.26.2681}}.
\newline\urlprefix\url{https://link.aps.org/doi/10.1103/PhysRevD.26.2681}

\bibitem{BlancoPillado2004}
J.~J. Blanco-Pillado, M.~Bucher, S.~Ghassemi, F.~Glanois,
  \href{https://link.aps.org/doi/10.1103/PhysRevD.69.103515}{When do colliding
  bubbles produce an expanding universe?}, Phys. Rev. D 69 (2004) 103515.
\newblock \href {https://doi.org/10.1103/PhysRevD.69.103515}
  {\path{doi:10.1103/PhysRevD.69.103515}}.
\newline\urlprefix\url{https://link.aps.org/doi/10.1103/PhysRevD.69.103515}

\bibitem{Freivogel2007}
B.~Freivogel, G.~T. Horowitz, S.~Shenker, {Colliding with a crunching bubble},
  JHEP 05 (2007) 090.
\newblock \href {http://arxiv.org/abs/hep-th/0703146}
  {\path{arXiv:hep-th/0703146}}, \href
  {https://doi.org/10.1088/1126-6708/2007/05/090}
  {\path{doi:10.1088/1126-6708/2007/05/090}}.

\bibitem{Easther2009}
R.~Easther, J.~T. Giblin, Jr, L.~Hui, E.~A. Lim, {A new mechanism for bubble
  nucleation: classical transitions}, Phys. Rev. D 80 (2009) 123519.
\newblock \href {http://arxiv.org/abs/0907.3234} {\path{arXiv:0907.3234}},
  \href {https://doi.org/10.1103/PhysRevD.80.123519}
  {\path{doi:10.1103/PhysRevD.80.123519}}.

\bibitem{Giblin2010}
J.~T. Giblin, Jr, L.~Hui, E.~A. Lim, I.-S. Yang, {How to run through walls:
  dynamics of bubble and soliton collisions}, Phys. Rev. D 82 (2010) 045019.
\newblock \href {http://arxiv.org/abs/1005.3493} {\path{arXiv:1005.3493}},
  \href {https://doi.org/10.1103/PhysRevD.82.045019}
  {\path{doi:10.1103/PhysRevD.82.045019}}.

\bibitem{Johnson2012}
M.~C. Johnson, H.~V. Peiris, L.~Lehner, {Determining the outcome of cosmic
  bubble collisions in full general relativity}, Phys. Rev. D 85 (2012) 083516.
\newblock \href {http://arxiv.org/abs/1112.4487} {\path{arXiv:1112.4487}},
  \href {https://doi.org/10.1103/PhysRevD.85.083516}
  {\path{doi:10.1103/PhysRevD.85.083516}}.

\bibitem{Hwang2012}
D.-i. Hwang, B.-H. Lee, W.~Lee, D.-h. Yeom, {Bubble collision with
  gravitation}, JCAP 07 (2012) 003.
\newblock \href {http://arxiv.org/abs/1201.6109} {\path{arXiv:1201.6109}},
  \href {https://doi.org/10.1088/1475-7516/2012/07/003}
  {\path{doi:10.1088/1475-7516/2012/07/003}}.

\bibitem{Bond2015}
J.~R. Bond, J.~Braden, L.~Mersini-Houghton, {Cosmic bubble and domain wall
  instabilities {III}: the role of oscillons in three-dimensional bubble
  collisions}, JCAP 09 (2015) 004.
\newblock \href {http://arxiv.org/abs/1505.02162} {\path{arXiv:1505.02162}},
  \href {https://doi.org/10.1088/1475-7516/2015/09/004}
  {\path{doi:10.1088/1475-7516/2015/09/004}}.

\bibitem{Aurrekoetxea2025}
J.~C. Aurrekoetxea, K.~Clough, E.~A. Lim, {Cosmology using numerical
  relativity}, Living Rev. Rel. 28~(1) (2025) 5.
\newblock \href {http://arxiv.org/abs/2409.01939} {\path{arXiv:2409.01939}},
  \href {https://doi.org/10.1007/s41114-025-00058-z}
  {\path{doi:10.1007/s41114-025-00058-z}}.

\bibitem{Takamizu2006}
Y.-i. Takamizu, K.-i. Maeda, {Collision of domain walls in asymptotically anti
  de Sitter spacetime}, Phys. Rev. D 73 (2006) 103508.
\newblock \href {http://arxiv.org/abs/hep-th/0603076}
  {\path{arXiv:hep-th/0603076}}, \href
  {https://doi.org/10.1103/PhysRevD.73.103508}
  {\path{doi:10.1103/PhysRevD.73.103508}}.

\bibitem{Takamizu2007}
Y.-i. Takamizu, H.~Kudoh, K.-i. Maeda, {Dynamics of colliding branes and black
  brane production}, Phys. Rev. D 75 (2007) 061304.
\newblock \href {http://arxiv.org/abs/gr-qc/0702138}
  {\path{arXiv:gr-qc/0702138}}, \href
  {https://doi.org/10.1103/PhysRevD.75.061304}
  {\path{doi:10.1103/PhysRevD.75.061304}}.

\bibitem{Omotani2011}
J.~Omotani, P.~M. Saffin, J.~Louko, {Colliding branes and big crunches}, Phys.
  Rev. D 84 (2011) 063526.
\newblock \href {http://arxiv.org/abs/1107.3938} {\path{arXiv:1107.3938}},
  \href {https://doi.org/10.1103/PhysRevD.84.063526}
  {\path{doi:10.1103/PhysRevD.84.063526}}.

\bibitem{Maeda2012}
K.-i. Maeda, K.~Uzawa, {Dynamical brane with angles: collision of the
  universes}, Phys. Rev. D 85 (2012) 086004.
\newblock \href {http://arxiv.org/abs/1201.3213} {\path{arXiv:1201.3213}},
  \href {https://doi.org/10.1103/PhysRevD.85.086004}
  {\path{doi:10.1103/PhysRevD.85.086004}}.

\bibitem{Tziolas2008}
A.~Tziolas, A.~Wang, {Colliding branes and formation of spacetime
  singularities}, Phys. Lett. B 661 (2008) 5--10.
\newblock \href {http://arxiv.org/abs/0704.1311} {\path{arXiv:0704.1311}},
  \href {https://doi.org/10.1016/j.physletb.2008.01.058}
  {\path{doi:10.1016/j.physletb.2008.01.058}}.

\bibitem{Tziolas2009}
A.~Tziolas, A.~Wang, Z.~C. Wu, {Colliding branes and formation of spacetime
  singularities in string theory}, JHEP 04 (2009) 038.
\newblock \href {http://arxiv.org/abs/0812.1377} {\path{arXiv:0812.1377}},
  \href {https://doi.org/10.1088/1126-6708/2009/04/038}
  {\path{doi:10.1088/1126-6708/2009/04/038}}.

\bibitem{Wainwright2014}
C.~L. Wainwright, M.~C. Johnson, A.~Aguirre, H.~V. Peiris, {Simulating the
  universe(s) {II}: phenomenology of cosmic bubble collisions in full general
  relativity}, JCAP 10 (2014) 024.
\newblock \href {http://arxiv.org/abs/1407.2950} {\path{arXiv:1407.2950}},
  \href {https://doi.org/10.1088/1475-7516/2014/10/024}
  {\path{doi:10.1088/1475-7516/2014/10/024}}.

\bibitem{Wainwright2014a}
C.~L. Wainwright, M.~C. Johnson, H.~V. Peiris, A.~Aguirre, L.~Lehner, S.~L.
  Liebling, {Simulating the universe(s): from cosmic bubble collisions to
  cosmological observables with numerical relativity}, JCAP 03 (2014) 030.
\newblock \href {http://arxiv.org/abs/1312.1357} {\path{arXiv:1312.1357}},
  \href {https://doi.org/10.1088/1475-7516/2014/03/030}
  {\path{doi:10.1088/1475-7516/2014/03/030}}.

\bibitem{Johnson2016}
M.~C. Johnson, C.~L. Wainwright, A.~Aguirre, H.~V. Peiris, {Simulating the
  universe(s) {III}: observables for the full bubble collision spacetime}, JCAP
  07 (2016) 020.
\newblock \href {http://arxiv.org/abs/1508.03641} {\path{arXiv:1508.03641}},
  \href {https://doi.org/10.1088/1475-7516/2016/07/020}
  {\path{doi:10.1088/1475-7516/2016/07/020}}.

\bibitem{Johnson2016a}
M.~C. Johnson, W.~Lin, {Observable signatures of a classical transition}, JCAP
  03 (2016) 051.
\newblock \href {http://arxiv.org/abs/1508.03786} {\path{arXiv:1508.03786}},
  \href {https://doi.org/10.1088/1475-7516/2016/03/051}
  {\path{doi:10.1088/1475-7516/2016/03/051}}.

\bibitem{Kim2015}
D.-H. Kim, B.-H. Lee, W.~Lee, J.~Yang, D.-h. Yeom, {Gravitational waves from
  cosmic bubble collisions}, Eur. Phys. J. C 75~(3) (2015) 133.
\newblock \href {http://arxiv.org/abs/1410.4648} {\path{arXiv:1410.4648}},
  \href {https://doi.org/10.1140/epjc/s10052-015-3348-2}
  {\path{doi:10.1140/epjc/s10052-015-3348-2}}.

\bibitem{Henneaux1985}
M.~Henneaux, {Quantum gravity in two dimensions: exact solution of the {Jackiw}
  model}, Phys. Rev. Lett. 54 (1985) 959--962.
\newblock \href {https://doi.org/10.1103/PhysRevLett.54.959}
  {\path{doi:10.1103/PhysRevLett.54.959}}.

\bibitem{Alwis1992}
S.~P. de~Alwis, {Quantization of a theory of 2-d dilaton gravity}, Phys. Lett.
  B 289 (1992) 278--282.
\newblock \href {http://arxiv.org/abs/hep-th/9205069}
  {\path{arXiv:hep-th/9205069}}, \href
  {https://doi.org/10.1016/0370-2693(92)91219-Y}
  {\path{doi:10.1016/0370-2693(92)91219-Y}}.

\bibitem{Vaz1994}
C.~Vaz, L.~Witten, {Formation and evaporation of a naked singularity in 2-d
  gravity}, Phys. Lett. B 325 (1994) 27--32.
\newblock \href {http://arxiv.org/abs/hep-th/9311133}
  {\path{arXiv:hep-th/9311133}}, \href
  {https://doi.org/10.1016/0370-2693(94)90066-3}
  {\path{doi:10.1016/0370-2693(94)90066-3}}.

\bibitem{Callan1992}
C.~G. Callan, Jr., S.~B. Giddings, J.~A. Harvey, A.~Strominger, {Evanescent
  black holes}, Phys. Rev. D 45~(4) (1992) R1005.
\newblock \href {http://arxiv.org/abs/hep-th/9111056}
  {\path{arXiv:hep-th/9111056}}, \href
  {https://doi.org/10.1103/PhysRevD.45.R1005}
  {\path{doi:10.1103/PhysRevD.45.R1005}}.

\bibitem{Bilal1993}
A.~Bilal, C.~G. Callan, Jr., {Liouville models of black hole evaporation},
  Nucl. Phys. B 394 (1993) 73--100.
\newblock \href {http://arxiv.org/abs/hep-th/9205089}
  {\path{arXiv:hep-th/9205089}}, \href
  {https://doi.org/10.1016/0550-3213(93)90102-U}
  {\path{doi:10.1016/0550-3213(93)90102-U}}.

\bibitem{Russo:1992ht}
J.~G. Russo, L.~Susskind, L.~Thorlacius, {Black hole evaporation in
  (1+1)-dimensions}, Phys. Lett. B 292 (1992) 13--18.
\newblock \href {http://arxiv.org/abs/hep-th/9201074}
  {\path{arXiv:hep-th/9201074}}, \href
  {https://doi.org/10.1016/0370-2693(92)90601-Y}
  {\path{doi:10.1016/0370-2693(92)90601-Y}}.

\bibitem{Russo:1992ax}
J.~G. Russo, L.~Susskind, L.~Thorlacius, {The endpoint of {Hawking} radiation},
  Phys. Rev. D 46 (1992) 3444--3449.
\newblock \href {http://arxiv.org/abs/hep-th/9206070}
  {\path{arXiv:hep-th/9206070}}, \href
  {https://doi.org/10.1103/PhysRevD.46.3444}
  {\path{doi:10.1103/PhysRevD.46.3444}}.

\bibitem{Thorlacius:1994ip}
L.~Thorlacius, {Black hole evolution}, Nucl. Phys. B Proc. Suppl. 41 (1995)
  245--275.
\newblock \href {http://arxiv.org/abs/hep-th/9411020}
  {\path{arXiv:hep-th/9411020}}, \href
  {https://doi.org/10.1016/0920-5632(95)00435-C}
  {\path{doi:10.1016/0920-5632(95)00435-C}}.

\bibitem{doi:10.1142/0622}
J.~D. Brown, \href{https://www.worldscientific.com/doi/abs/10.1142/0622}{Lower
  dimensional gravity}, World Scientific, 1988.
\newblock \href
  {https://doi.org/10.1142/0622} {\path{doi:10.1142/0622}}.
\newline\urlprefix\url{https://www.worldscientific.com/doi/abs/10.1142/0622}

\bibitem{Nojiri2001}
S.~Nojiri, S.~D. Odintsov, {Quantum dilatonic gravity in $d=2$, 4 and 5
  dimensions}, Int. J. Mod. Phys. A 16 (2001) 1015--1108.
\newblock \href {http://arxiv.org/abs/hep-th/0009202}
  {\path{arXiv:hep-th/0009202}}, \href
  {https://doi.org/10.1142/S0217751X01002968}
  {\path{doi:10.1142/S0217751X01002968}}.

\bibitem{Grumiller:2002nm}
D.~Grumiller, W.~Kummer, D.~V. Vassilevich, {Dilaton gravity in
  two-dimensions}, Phys. Rept. 369 (2002) 327--430.
\newblock \href {http://arxiv.org/abs/hep-th/0204253}
  {\path{arXiv:hep-th/0204253}}, \href
  {https://doi.org/10.1016/S0370-1573(02)00267-3}
  {\path{doi:10.1016/S0370-1573(02)00267-3}}.

\bibitem{Mertens:2022irh}
T.~G. Mertens, G.~J. Turiaci, {Solvable models of quantum black holes: a review
  on Jackiw{\textendash}Teitelboim gravity}, Living Rev. Rel. 26~(1) (2023) 4.
\newblock \href {http://arxiv.org/abs/2210.10846} {\path{arXiv:2210.10846}},
  \href {https://doi.org/10.1007/s41114-023-00046-1}
  {\path{doi:10.1007/s41114-023-00046-1}}.

\bibitem{Mann1991}
R.~B. Mann, S.~M. Morsink, A.~E. Sikkema, T.~G. Steele, {Semiclassical gravity
  in $(1+1)$ dimensions}, Phys. Rev. D 43 (1991) 3948--3957.
\newblock \href {https://doi.org/10.1103/PhysRevD.43.3948}
  {\path{doi:10.1103/PhysRevD.43.3948}}.

\bibitem{Stoetzel1995}
B.~St\"otzel, {Two-dimensional gravitation and sine-Gordon solitons}, Phys. Rev.
  D 52 (1995) 2192--2201.
\newblock \href {http://arxiv.org/abs/gr-qc/9501033}
  {\path{arXiv:gr-qc/9501033}}, \href
  {https://doi.org/10.1103/PhysRevD.52.2192}
  {\path{doi:10.1103/PhysRevD.52.2192}}.

\bibitem{AlonsoIzquierdo:2019hvl}
A.~Alonso~Izquierdo, W.~Garc{\'\i}a~Fuertes, J.~Mateos~Guilarte,
  {Self-gravitating kinks in two-dimensional pseudo-Riemannian universes},
  Phys. Rev. D 101~(3) (2020) 036020.
\newblock \href {http://arxiv.org/abs/1911.08167} {\path{arXiv:1911.08167}},
  \href {https://doi.org/10.1103/PhysRevD.101.036020}
  {\path{doi:10.1103/PhysRevD.101.036020}}.

\bibitem{Zhong2021}
Y.~Zhong, {Revisit on two-dimensional self-gravitating kinks: superpotential
  formalism and linear stability}, JHEP 04 (2021) 118.
\newblock \href {http://arxiv.org/abs/2101.10928} {\path{arXiv:2101.10928}},
  \href {https://doi.org/10.1007/JHEP04(2021)118}
  {\path{doi:10.1007/JHEP04(2021)118}}.

\bibitem{Zhong:2021voa}
Y.~Zhong, F.-Y. Li, X.-D. Liu, {K-field kinks in two-dimensional dilaton
  gravity}, Phys. Lett. B 822 (2021) 136716.
\newblock \href {http://arxiv.org/abs/2108.10166} {\path{arXiv:2108.10166}},
  \href {https://doi.org/10.1016/j.physletb.2021.136716}
  {\path{doi:10.1016/j.physletb.2021.136716}}.

\bibitem{Zhong:2022pkv}
Y.~Zhong, {Singular {P{\"o}schl-Teller} {II} potentials and gravitating kinks},
  JHEP 09 (2022) 165.
\newblock \href {http://arxiv.org/abs/2207.12681} {\path{arXiv:2207.12681}},
  \href {https://doi.org/10.1007/JHEP09(2022)165}
  {\path{doi:10.1007/JHEP09(2022)165}}.

\bibitem{Feng:2022ndn}
J.~Feng, Y.~Zhong, {Scalar perturbation of gravitating double-kink solutions},
  EPL 137~(4) (2022) 49001.
\newblock \href {http://arxiv.org/abs/2202.02946} {\path{arXiv:2202.02946}},
  \href {https://doi.org/10.1209/0295-5075/ac56ae}
  {\path{doi:10.1209/0295-5075/ac56ae}}.

\bibitem{Zhong:2023pel}
Y.~Zhong, H.~Guo, Y.-X. Liu, {Kink solutions in generalized 2D dilaton
  gravity}, Phys. Lett. B 849 (2024) 138471.
\newblock \href {http://arxiv.org/abs/2308.13786} {\path{arXiv:2308.13786}},
  \href {https://doi.org/10.1016/j.physletb.2024.138471}
  {\path{doi:10.1016/j.physletb.2024.138471}}.

\bibitem{Wang:2024dbf}
H.~Wang, Y.~Zhong, Z.~Wang, {Rosen-{Morse} potential and gravitating kinks},
  Phys. Lett. B 858 (2024) 139071.
\newblock \href {http://arxiv.org/abs/2409.14761} {\path{arXiv:2409.14761}},
  \href {https://doi.org/10.1016/j.physletb.2024.139071}
  {\path{doi:10.1016/j.physletb.2024.139071}}.

\bibitem{Wang:2024kxo}
Z.~Wang, Y.~Zhong, H.~Wang, {Gravitating kinks with asymptotically flat
  metrics}, EPL 146~(5) (2024) 59001.
\newblock \href {http://arxiv.org/abs/2402.05486} {\path{arXiv:2402.05486}},
  \href {https://doi.org/10.1209/0295-5075/ad49d0}
  {\path{doi:10.1209/0295-5075/ad49d0}}.

\bibitem{Vachaspati2016}
T.~Vachaspati, {Creation of magnetic monopoles in classical scattering}, Phys.
  Rev. Lett. 117 (2016) 181601.
\newblock \href {http://arxiv.org/abs/1607.07460}
  {\path{arXiv:1607.07460}}, \href
  {https://doi.org/10.1103/PhysRevLett.117.181601}
  {\path{doi:10.1103/PhysRevLett.117.181601}}.

\bibitem{Zenginoglu2024}
A.~Zengino{\u{g}}lu, {Hyperbolic times in {Minkowski} space}, American Journal
  of Physics 92~(12) (2024) 965--974.
\newblock \href {http://arxiv.org/abs/2404.01528} {\path{arXiv:2404.01528}},
  \href {https://doi.org/10.1119/5.0214271} {\path{doi:10.1119/5.0214271}}.

\bibitem{Friedrich:1998xmu}
H.~Friedrich, {Gravitational fields near space-like and null infinity}, J.
  Geom. Phys. 24~(2) (1998) 83--163.
\newblock \href {https://doi.org/10.1016/s0393-0440(97)82168-7}
  {\path{doi:10.1016/s0393-0440(97)82168-7}}.

\bibitem{Evslin2021}
J.~Evslin, {Evidence for the unbinding of the $\phi^4$ kink's shape mode}, JHEP
  09 (2021) 009.
\newblock \href {http://arxiv.org/abs/2104.14387} {\path{arXiv:2104.14387}},
  \href {https://doi.org/10.1007/JHEP09(2021)009}
  {\path{doi:10.1007/JHEP09(2021)009}}.

\bibitem{Evslin2022}
J.~Evslin, A.~Garc{\'\i}a Mart{\'\i}n-Caro, {Spontaneous emission from excited
  quantum kinks}, JHEP 12 (2022) 111.
\newblock \href {http://arxiv.org/abs/2210.13791} {\path{arXiv:2210.13791}},
  \href {https://doi.org/10.1007/JHEP12(2022)111}
  {\path{doi:10.1007/JHEP12(2022)111}}.

\bibitem{Evslin2023}
J.~Evslin, H.~Liu, {(Anti-)Stokes scattering on kinks}, JHEP 03 (2023) 095.
\newblock \href {http://arxiv.org/abs/2301.04099} {\path{arXiv:2301.04099}},
  \href {https://doi.org/10.1007/JHEP03(2023)095}
  {\path{doi:10.1007/JHEP03(2023)095}}.

\bibitem{Evslin2024}
J.~Evslin, H.~Liu, {Elastic kink-meson scattering}, JHEP 04 (2024) 072.
\newblock \href {http://arxiv.org/abs/2311.14369} {\path{arXiv:2311.14369}},
  \href {https://doi.org/10.1007/JHEP04(2024)072}
  {\path{doi:10.1007/JHEP04(2024)072}}.

\bibitem{Bayarsaikhan2026}
B.~Bayarsaikhan, J.~Evslin, {A resonance in elastic kink-meson scattering}
\newblock \href {http://arxiv.org/abs/2603.26070} {\path{arXiv:2603.26070}}.

\end{thebibliography}

\end{document}